\def\j{{\bf j}}
\def\x{{\bf x}}
\def\p{{\bf p}}
\def\q{{\bf q}}
\def\k{{\bf k}}
\def\v{{\bf v}}
\def\u{{\bf u}}
\def\E{{\bf E}}

\def\magp{|{\bf p}|}
\def\f{{\overline{f}}}

\def\A{{\rm A}}
\def\B{{\rm B}}
\def\V{{\rm V}}

\def\LHS{{\rm LHS}}
\def\naBla{{\bf \nabla}}
\def\ij{{i \cdots j}}

\def\ca{C_{\rm A}}
\def\cf{C_{\rm F}}
\def\da{d_{\rm A}}
\def\ta{T_{\rm A}}
\def\tf{T_{\rm F}}
\def\df{d_{\rm F}}
\def\nf{N_{\rm f}\,}

\def\trans{\top}
\def\suscept{\Xi}
\newcommand\ansatz{{ansatz}}
\newcommand\ansatze{{ans\"{a}tze}}

\def\grad{\mbox{\boldmath$\nabla$}}
\def\half{{\textstyle{1\over2}}}

\def\gs{g_{\rm s}}
\def\gw{g_{\rm w}}
\def\alphaw{\alpha_{\rm w}}
\def\alphas{\alpha_{\rm s}}
\def\alphaEM{\alpha_{\rm EM}}

\def\chie{\chi^{e^{+}\!}}

\def\gsim{\mbox{~{\raisebox{0.4ex}{$>$}}\hspace{-1.1em}
	{\raisebox{-0.6ex}{$\sim$}}~}}
\def\lsim{\mbox{~{\raisebox{0.4ex}{$<$}}\hspace{-1.1em}
	{\raisebox{-0.6ex}{$\sim$}}~}}

\def\diffC{A}
\def\ffh{{\rm f\bar fh}}
\def\ffhc{{\rm f\bar fhc}}
\def\fh{{\rm fh}}

\newfam\scrfam                                          
\global\font\twelvescr=rsfs10 scaled\magstep1%
\global\font\eightscr=rsfs7 scaled\magstep1%
\global\font\sixscr=rsfs5 scaled\magstep1%
\skewchar\twelvescr='177\skewchar\eightscr='177\skewchar\sixscr='177%
\textfont\scrfam=\twelvescr\scriptfont\scrfam=\eightscr
\scriptscriptfont\scrfam=\sixscr
\def\scr{\fam\scrfam}%

\documentstyle[preprint,aps,fixes,epsf,eqsecnum]{revtex}

\makeatletter           
\@floatstrue
\def\figure{\let\@capwidth\columnwidth\@float{figure}}
\let\endfigure\end@float
\@namedef{figure*}{\let\@capwidth\textwidth\@dblfloat{figure}}
\@namedef{endfigure*}{\end@dblfloat}
\def\table{\let\@capwidth\columnwidth\@float{table}}
\let\endtable\end@float
\@namedef{table*}{\let\@capwidth\textwidth\@dblfloat{table}}
\@namedef{endtable*}{\end@dblfloat}
\def\la{\label}
\makeatother

\tightenlines
\newcommand\pcite[1]{\protect{\cite{#1}}}

\begin {document}


\preprint {UW/PT 00--15}

\title
    {
    Transport coefficients in high temperature gauge theories:\\
    (I) Leading-log results
    }

\author {Peter Arnold}
\address
    {%
    Department of Physics,
    University of Virginia,
    Charlottesville, VA 22901
    }%
\author{Guy D. Moore and Laurence G. Yaffe}
\address
    {%
    Department of Physics,
    University of Washington,
    Seattle, Washington 98195
    }%

\date {October 2000}

\maketitle
\vskip -20pt

\begin {abstract}%
    {%
    Leading-log results are derived for the shear viscosity,
    electrical conductivity, and flavor diffusion constants
    in both Abelian and non-Abelian high temperature gauge theories
    with various matter field content.
    }%
\end {abstract}

\thispagestyle{empty}

\section {Introduction and Summary}
\la{sec:intro}

    Transport coefficients, such as viscosities, diffusivities, or
conductivity, characterize the dynamics of long wavelength,
low frequency fluctuations in a medium.
In condensed matter applications transport coefficients are
typically measured, not calculated from first principles,
due to the complexity of the underlying microscopic dynamics.
But in a weakly coupled quantum field theory,
transport coefficients should, in principle, be calculable 
purely theoretically.
Knowledge of various transport coefficients in high temperature
gauge theories is important in cosmological applications
such as electroweak baryogenesis
\cite {baryogenesis1,baryogenesis2},
as well as hydrodynamic models of heavy ion collisions \cite {heavy-ion}.

    In this paper, we consider the evaluation of transport coefficients
in weakly coupled high temperature gauge theories.
``High temperature'' is taken to mean that the temperature
is much larger than the zero-temperature masses of elementary
particles, and any chemical potentials.
In QED, this means $T \gg m_e$, while in
QCD, we require both $T \gg \Lambda_{\rm QCD}$ and $T \gg m_q$.
Corrections suppressed by powers of temperature
($m_q/T$, $\Lambda_{\rm QCD}/T$, {\em etc}.) will be ignored.
This means that each transport coefficient will equal
some power of temperature, trivially determined by dimensional analysis,
multiplied by some function of the dimensionless coupling constants of the theory.

As an example, the shear viscosity in a (single component, real)
${\lambda \over 4!} \phi^4$ scalar theory has the high temperature form
\begin {equation}
    \eta = a \, {T^3 \over \lambda^2} \,,
\end {equation}
with $a = 3033.5$, up to relative corrections
suppressed by higher powers of $\lambda$ \cite {JeonYaffe,Jeon}.%
\footnote
    {
    The value quoted for $a$ actually comes from our own evaluation
    of the $\phi^4$ shear viscosity, using the
    variational formulation described in section \ref {sec:kinetic}.
    This allows a
    higher precision evaluation than that obtained by discretizing
    the requisite integral equation as described in \cite {Jeon}.
    }
The leading $1/\lambda^{2}$ behavior reflects the fact that
the two-body scattering cross section is $O(\lambda^2)$,
and that transport coefficients are inversely proportional
to scattering rates.

In gauge theories, the presence of Coulomb scattering 
over a parametrically wide range of momentum transfers
(or scattering angles) causes transport coefficients
to have a more complicated dependence on the interaction strength.
In QED, for example, the high temperature shear viscosity has the form
\begin {equation}
    \eta = \kappa \, {T^3 \over \alpha^2 \, \ln \alpha^{-1}} \,,
\end {equation}
up to relative corrections which are
suppressed by additional factors of $1/\ln \alpha^{-1}$.
Evaluating the overall constant $\kappa$, while ignoring
all terms suppressed by additional powers of $1/\ln \alpha^{-1}$
(or powers of $\alpha$)
amounts to a ``leading-log'' calculation of the transport coefficient.

In this paper, we will present leading-log calculations of
the shear viscosity, electrical conductivity, and flavor
diffusion constants in high temperature gauge theories
(Abelian or non-Abelian) with various matter field content.%
\footnote
    {%
    We will not consider the bulk viscosity.
    It requires a significantly different, and more complicated,
    analysis than other transport coefficients.
    We also do not treat thermal conductivity, which is not an 
    independent transport coefficient in the absence of nonzero 
    conserved charges (besides energy and momentum).
    }
It should be emphasized, however, that these leading-log results
cannot be presumed to provide a quantitatively reliable determination
of transport coefficients in any real application.
Gauge couplings in the standard model are never so tiny
that corrections suppressed by $1/\ln \alphas^{-1}$,
$1/\ln \alphaw^{-1}$, or even $1/\ln \alphaEM^{-1}$ are negligibly small.
Nevertheless, the leading-log analysis of transport coefficients
is a useful first step.
In a companion paper, we extend our treatment and obtain
``all-log'' results which include all terms suppressed only
by inverse logarithms of the gauge coupling (but drop sub-leading
effects suppressed by powers of the coupling) \cite {all-log}.

Previous efforts to determine transport coefficients in hot
gauge theories include many applications of {\it relaxation time}
approximations
\cite {HosoyaKajantie,Hosoya_and_co,relax1,relax2,relax3,relax4,bad1,bad2,bad3,bad4},
in which the full momentum dependence of
relevant scattering rates are crudely characterized by a single
relaxation time.
Such treatments can, at best, obtain the correct leading parametric dependence
on the coupling and a rough estimate of the overall coefficient (though
some of them \cite{bad1,bad2,bad3,bad4} do not obtain the right parametric
behavior).  In addition, there have been a number of papers
reporting genuine leading-log evaluations of various transport coefficients
\cite {BMPRa,heiselberg,%
	Heiselberg_diff,BaymHeiselberg,JPT1,MooreProkopec,JPT2}.
However, we find that almost all of these results are
incorrect due to a variety of both conceptual and technical errors.
(In some cases \cite{BMPRa}, the errors are numerically quite small.)
For each transport coefficient we consider,
specific comparisons with previous work will be detailed in
the relevant section below.

In section \ref {sec:kinetic} we discuss how one may construct a
linearized kinetic theory which is adequate for computing correctly the
transport coefficients we consider, up to corrections
suppressed by powers of coupling.
We also show how the actual calculation of a transport coefficient
may be converted into a variational problem;
this is very convenient for numerical purposes.
(Related, but somewhat different variational formulations
appear in the literature.)
For leading-log calculations one may greatly simplify the resulting
collision integrals, since the coefficient of the leading-log is
only sensitive to small angle scattering processes.
This is discussed in section \ref {sec:leading-log}.
The details of the analysis for the electrical conductivity,
flavor diffusivities, and shear viscosity are presented
in sections \ref {sec:conductivity}, \ref {sec:diffusion},
and \ref {sec:shear}, respectively.
Throughout this paper, we will present results for
arbitrary simple gauge groups, rather than specializing to SU($N$).
Our notation for group factors ($C_{\rm R}$, $T_{\rm R}$, $d_{\rm R}$),
their SU($N$) values, and how to interpret them for Abelian problems,
is explained in Appendix \ref{app:group}.

In the remainder of this introduction, we review the basic
definitions of the various transport coefficients,
and then summarize our results.

\subsection {Definitions}

    At sufficiently high temperature, the equilibrium state of any
relativistic field theory may be regarded as a relativistic fluid.
The stress energy tensor $T_{\mu\nu}$ defines four locally-conserved
currents whose corresponding conserved charges are, of course,
the total energy and spatial momentum of the system.
At any point in the system,
the local fluid rest frame is defined as the frame in which
the local momentum density vanishes,%
\footnote
    {%
    This is often termed the Landau-Lifshitz convention.
    In theories with additional conserved currents,
    such as a baryon number current, one may alternatively
    define the local rest frame as the frame in which 
    there is no baryon number flux.
    This is the Eckart convention.
    As we wish to consider, among others, pure gauge theories
    in which no other conserved currents are present,
    the Landau-Lifshitz convention provides the only uniform
    definition of local flow.
    }
\begin {equation}
    T^0_{\;\;i}(x) = 0 \quad \Leftrightarrow \quad
    \hbox {[local fluid rest frame at $x$]} \,.
\end {equation}

If the fluid is slightly disturbed from equilibrium,
then the non-equilibrium expectation of $T_{\mu\nu}$,
in the local fluid rest frame,
satisfies the constitutive relation%
\footnote
    {%
    We use $({-}{+}{+}{+})$ metric conventions.
    }
\begin {equation}
    \langle T_{ij} \rangle
    =
    \delta_{ij} \, \langle {\scr P} \rangle
    - \eta \left[ \nabla_i \, u_j + \nabla_j \, u_i
		- {\textstyle {2\over3}} \, \delta_{ij} \, \nabla^l \, u_l
	    \right]
    - \zeta \, \delta_{ij} \, \nabla^l \, u_l \,,
\la {eq:Tij}
\end {equation}
together with the exact conservation law
$
    \partial_\mu \langle T^{\mu\nu} \rangle = 0
$.
In the constitutive relation (\ref {eq:Tij}),
$\eta$ is the {\em shear viscosity},
$\zeta$ is the {\em bulk viscosity},
and $\u$ is local flow velocity.
For small departures from equilibrium,
\begin {equation}
    u_i \equiv
    \langle T^0_{\;\;i} \rangle / \langle \varepsilon + {\scr P} \rangle \,,
\end {equation}
where $\varepsilon \equiv T^{00}$ is the energy density
and ${\scr P} \equiv {1\over3} T^{i}_{\;\;i}$ the (local) pressure.
The combination $\varepsilon + {\scr P}$ is also known as the enthalpy.
The constitutive relation (\ref {eq:Tij}) holds up to
corrections involving further gradients or higher powers of $\u$.

In a similar fashion,
in any theory containing electromagnetism
({\em i.e.}, a $U(1)$ gauge field), the electric current
density $j^{\rm EM}_\mu$ is conserved ($\partial^\mu j^{\rm EM}_\mu = 0$)
and satisfies the constitutive relation
\begin {equation}
    \langle j^{\rm EM}_i \rangle = \sigma \, \langle E_i \rangle \,,
\la {eq:jiEM}
\end {equation}
for small departures from equilibrium.
Here $\sigma$ is the (DC) {\em electrical conductivity},

And in theories containing one or more conserved ``flavor'' currents,
$ j^\alpha_\mu $,
which are not coupled to dynamical gauge fields
(such as baryon number or isospin currents in QCD),
these currents will satisfy diffusive constitutive relations,
\begin {equation}
    \langle j^\alpha_i \rangle
    = -D^{\alpha\beta} \> \nabla_i \langle n^\beta \rangle \,,
\la {eq:jia}
\end {equation}
where $n^\alpha \equiv (j^\alpha)^0$ is a conserved charge density,
and $D \equiv ||D^{\alpha\beta}||$ is, in general, a matrix of
{\em diffusion constants}.
Once again, the constitutive relation (\ref {eq:jia}) holds in the
local fluid rest frame; otherwise, an additional convective
$\langle n^\alpha \rangle \, v_i$ term is present.

In the limit of arbitrarily small gradients
(so that the scale of variation in
$\langle T_{\mu\nu} \rangle$ or $\langle j_\mu \rangle$
is huge compared to microscopic length scales)
and arbitrarily small departures from equilibrium,
the constitutive relations (\ref {eq:Tij}), (\ref {eq:jiEM}),
and (\ref {eq:jia})
may be regarded as definitions of the shear and bulk viscosities,
electrical conductivity, and flavor diffusion constants.
Alternatively, one may use linear response theory to relate
non-equilibrium expectation values to equilibrium correlation functions
\cite {Hosoya_and_co}.
This leads to well-known Kubo relations which express transport
coefficients in terms of the zero-frequency slope of spectral densities
of current-current, or stress tensor--stress tensor correlation functions,
\begin {mathletters}
\begin {eqnarray}
    \eta &=&
    {1 \over 20} \, \lim_{\omega \to 0} {1 \over \omega}
    \int d^4x \> e^{i \omega t} \>
    \langle [ \pi_{lm}(t,\x), \, \pi_{lm}(0) ] \rangle_{\rm eq} \,,
\la {eq:etaKubo}
\\
    \zeta &=&
    {1 \over 2} \, \lim_{\omega \to 0} {1 \over \omega}
    \int d^4x \> e^{i \omega t} \>
    \langle [ {\scr P}(t,\x), \, {\scr P}(0) ] \rangle_{\rm eq} \,,
\la {eq:zetaKubo}
\\
    \sigma &=&
    {1 \over 6} \, \lim_{\omega \to 0} {1 \over \omega}
    \int d^4x \> e^{i \omega t} \>
    \langle [ j_i^{\rm EM}(t,\x), \, j_i^{\rm EM}(0) ] \rangle_{\rm eq} \,,
\la {eq:sigmaKubo}
\\
    D_{\alpha\beta} &=&
    {1 \over 6} \, \lim_{\omega \to 0} {1 \over \omega}
    \int d^4x \> e^{i \omega t} \>
    \langle [ j_i^\alpha(t,\x), \, j_i^\gamma(0) ] \rangle_{\rm eq}
    \; \suscept^{-1}_{\gamma\beta}
    \,.
\la {eq:DKubo}
\end {eqnarray}%
\la {eq:Kubo}%
\end {mathletters}%
In Eq.~(\ref {eq:etaKubo}),
$\pi_{lm} \equiv T_{lm} - \delta_{lm} {\scr P}$
denotes the traceless part of the stress tensor,
while in Eq.~(\ref {eq:DKubo}),
$\suscept \equiv ||\suscept_{\alpha\beta}||$ is the
``charge susceptibility'' matrix
describing mean-square global charge fluctuations (per unit volume),
\begin {equation}
    \suscept_{\alpha\beta} \equiv
    {\partial \langle n^\alpha \rangle \over \partial\mu_\beta}
    =
    {\beta \over {\scr V}}
    \left[
	\langle N^\alpha N^\beta \rangle -
	\langle N^\alpha \rangle \langle N^\beta \rangle
    \right] ,
\la {eq:susceptability}
\end {equation}
where $N^\alpha \equiv \int (d^3\x) \> n^\alpha$ is a conserved charge,
$\scr V$ is the spatial volume, and $\beta$ is the inverse temperature.

\bigskip

\subsection {Results}

\subsubsection {Electrical conductivity}

The high temperature electrical conductivity has the leading-log form
\begin {equation}
    \sigma = C \, { T \over e^2 \, \ln e^{-1} } \,,
\end {equation}
where the dimensionless coefficient $C$ depends on the
number (and relative charges) of the electrically charged
matter fields which couple to the photon.
The $T/(e^2 \,\ln e^{-1})$ dependence of $\sigma$ can be
qualitatively understood as arising from the form
$\sigma \sim e^2 \, T^2 \, \tau$,
where $\tau \sim \left[ e^4 T \, \log e^{-1}\right]^{-1} $
is the characteristic time scale for large-angle scattering
(from either a single hard scattering or a sequence of
small-angle scatterings which add up to produce a large angle deflection),
or ``transport mean free time.''%
\footnote
    {%
    The transport mean free time is the inverse collision rate
    for large-angle scattering ({\em i.e.}, an $O(1)$ change in direction).
    In high temperature gauge theories, the mean free time for
    any scattering is dominated by very small angle scattering
    and is of order $(g^2 T)^{-1}$, up to logarithms,
    while the transport mean free time is order $(g^4 T \ln g^{-1})^{-1}$
    \cite {HosoyaKajantie,relax1,relax2,relax3,relax4,BMPRa,SelikhovGyulassy,smilga,A&Y}.%
    }
This is just the classic Drude model form
$\sigma \sim n \, e^2 \, \tau /m$,
appropriately generalized to an ultra-relativistic setting
(so $n \sim T^3$, and $m$ is replaced by the typical energy $T$).
Our exact results for the leading-log coefficient~$C$,
for various subsets of the fermions of the standard model,
are shown in table~\ref {cond_table}.  Which entry in the table to use
depends on the temperature; each entry is valid when the included
fermions are light ($m \ll T$) but the excluded fermions are heavy.

\begin {table}[t]
\begin{center}
\tabcolsep 10pt
\begin{tabular}{cccc}
species & $N_{\rm leptons}$ & $N_{\rm species}$
	& $\sigma \times (e^2/T) \ln e^{-1}$    \\\hline
$e$			& 1 & 1    & 15.6964 \\
$e,\mu$			& 2 & 2    & 20.6566 \\
$e,\mu,u,d,s$		& 2 & 4    & 12.2870 \\
$e,\mu,\tau,u,d,s,c$	& 3 & 19/3 & 12.5202 \\
$e,\mu,\tau,u,d,s,c,b$	& 3 & 20/3 & 11.9719
\end{tabular}%
\end{center}
\caption
    {%
    \label{cond_table}
    Leading-log conductivity for various numbers of
    leptonic charge carriers ($N_{\rm leptons}$)
    and effective number of leptonic plus quark scatterers
    ($N_{\rm species}$).
    The number $N_{\rm species}$ is a sum over all fermion
    fields weighted by the square of their electric charge.
    }
\end{table}

The dependence on matter field content is not precisely given by
any simple analytic formula.
But as detailed in section \ref {sec:conductivity},
the conductivity is approximately equal to
\begin {equation}
    \sigma \simeq
    \left(
	{12^4 \, \zeta(3)^2 \, \pi^{-3} \, N_{\rm leptons}
	\over 3 \pi^2 + 32 \, N_{\rm species}}
    \right)
    {T \over e^2 \ln e^{-1}} \,,
\la {eq:approx-cond}
\end {equation}
where $N_{\rm leptons}$ is the number of leptonic charge carriers
(not counting anti-particles separately from particles),
and $N_{\rm species}$ is a sum over all (Dirac) fermion fields weighted
by the square of their electric charge assignments.
So each leptonic species contributes 1 to $N_{\rm species}$,
each down type quark contributes 1/3 [which is $(-1/3)^2$ times 3 colors],
and each up type quark contributes 4/3.
It may be noted that this is exactly the same sum over charged species
which appears in the (lowest-order) expression for the high temperature
Debye screening mass for the photon,
\begin {equation}
    m_D^2 = {\textstyle {1 \over 3}} \, N_{\rm species} \, e^2 T^2 \,.
\la {eq:debye}
\end {equation}

The approximate form (\ref {eq:approx-cond})
is the result of a one-term variational approximation
which becomes exact if the correct quantum statistics is replaced by
classical Boltzmann statistics.
This approximation reproduces the true leading-log coefficient
to within an accuracy of better than 0.4\% for all cases shown
in Table \ref {cond_table}.
In practice,
neglect of subleading effects suppressed by powers of $1/\ln e^{-1}$
is a far larger issue than the accuracy with which Eq.~(\ref {eq:approx-cond})
reproduces the exact leading-log coefficient.

Our Eq.~(\ref{eq:approx-cond}) is similar to the expression found by Baym
and Heiselberg \cite{BaymHeiselberg}.  The most significant difference
is that their expression is missing the $3 \pi^2$ term in the
denominator.
This term arises from Compton scattering and annihilation to
photons --- processes neglected in Ref.~\cite {BaymHeiselberg}.

At temperatures above the QCD scale,
where scattering from quarks must be included,
these leading-log results
neglect the contribution of quarks
to the electric current density $j_\mu^{\rm EM}$ itself.
This is quite a good approximation since the rate of strong interactions
among quarks is much greater than their electromagnetic interactions,
and will wash out departures from equilibrium in quark distributions
much faster than the relaxation of fluctuations in lepton distributions,
which depends on electromagnetic interactions.  This simplification
amounts to the neglect of corrections to the coefficient $C$ suppressed 
by $\alpha_{\rm EM}^2 / \alphas^2$.  There are also relative
$O(\alpha_s)$ corrections arising from QCD effects when a lepton scatters
from a quark.

Weak interactions are ignored altogether in these calculations,
so the results above, when applied to the standard model,
are relevant at temperatures small compared to $M_W$, but large
compared to the masses of the quarks and leptons considered.
In particular, the electromagnetic mean free path
for large angle scattering must be small compared to the
weak interaction mean free path which, parametrically,
requires that $\alpha^2 \gg \alphaw^2 (T/M_W)^4$.

For temperatures comparable or large compared to $M_W$, where the
electroweak sector of the standard model is in its ``unbroken''
high temperature phase,%
\footnote
    {
    Depending on details of the scalar sector,
    there need not be any sharp electroweak phase transition.
    The ``high temperature electroweak phase'' should be understood as
    the regime where the effective mass of the weak gauge bosons
    comes predominantly from thermal fluctuations, not the Higgs condensate.
    In other words, $T \gsim v(T)$, where $v(T)$ is the
    (temperature dependent) Higgs expectation value.
    }
the dynamics of the $U(1)$ hypercharge field may be
characterized by a hypercharge conductivity
in complete analogy to ordinary electromagnetism.
Neglecting relative corrections suppressed by $\tan^4 \theta_{\rm W}$
and $\alpha'/\alphas$,
and negligible effects due to Yukawa couplings of right-handed leptons,
the hypercharge conductivity
is determined by the hypercharge-mediated scattering of right-handed leptons.
(Quarks and left-handed leptons scatter much more rapidly
due to $SU(2)$ or $SU(3)$ interactions, and hence
contribute much less to the conductivity than right-handed leptons.)
The appropriate generalization of Eq.~(\ref {eq:approx-cond})
has exactly the same form but with the charge $e$ replaced
by the hypercharge coupling $g'$,
$N_{\rm leptons}$ replaced by half the number of right-handed leptons,
and $N_{\rm species}$ now given by
the sum of the square of the hypercharge of each complex scalar field,
plus half the square of the hypercharge of each chiral fermion field.
For the three generations of the standard model, this means
$N_{\rm leptons} = 3/2$, and $N_{\rm species} = 5 + n_{\rm s}/2$,
where $n_{\rm s}$ is the number of Higgs doublets,%
\footnote
    {%
    We normalize ``hypercharge'' $Y$ by $Q = T_3 + Y$ (as opposed to the
    other convention that $Q = T_3 + Y/2$).
    The conductivity (\ref{eq:hypercond}) does not depend on this normalization
    convention.
    In greater detail,
    $
	N_{\rm species} 
	=
	2 n_{\rm s} (1/2)^2
	+
	3/2
	\left[
	    (-1)^2 + 2 (-1/2)^2 + 3 (2/3)^2 + 3 (-1/3)^2 + 6 (1/6)^2
	\right]
    $,
    where the various terms in the bracket come from
    right-handed leptons,
    left-handed leptons,
    right-handed up type quarks,
    right-handed down type quarks,
    and left-handed quarks, respectively.
    The Debye mass of the $U(1)$ hypercharge gauge field again satisfies
    Eq.~(\ref {eq:debye}), with $e^2 \to g'{}^2$.
    }
so that leading-log hypercharge conductivity is (approximately)
\begin {equation}
    \sigma_{\rm hyper} \simeq
    6^4 \, \zeta(3)^2 \, \pi^{-3}
    \left[
	{\pi^2 \over 8} + {20 \over 3} + {2 \over 3} n_{\rm s}
    \right]^{-1}
    \left({T \over g'{}^2 \ln g'{}^{-1}}\right) \,.
\la {eq:hypercond}
\end {equation}

\subsubsection {Shear viscosity}

The high temperature shear viscosity
in a gauge theory with a simple gauge group
(either Abelian or non-Abelian)
has the leading-log form
\begin {equation}
    \eta = \kappa \, {T^3 \over g^4 \, \ln g^{-1}} \,,
\end {equation}
where $g$ is the gauge coupling.
For the case of $SU(3)$ gauge theory ({\em i.e.}, QCD),
our results for the leading-log shear viscosity coefficient $\kappa$
for various numbers of fermion species
are shown in table \ref {shear_table}.

\begin{table}[tb]
\tabcolsep 10pt
\begin {center}
\begin{tabular}{cc}
$ \quad \nf \quad $ & $ \eta \times (g^4 / T^3) \ln g^{-1} $ \\ \hline 
0 & \phantom{1}27.126  \\
1 & \phantom{1}60.808  \\
2 & \phantom{1}86.473  \\
3 & 106.664  \\
4 & 122.958  \\
5 & 136.380  \\
6 & 147.627
\end{tabular}
\end {center}
\caption
    {%
    \label{shear_table}
    Leading-log shear viscosity as a function of the
    number of (fundamental representation) fermion flavors with $m \ll T$,
    for gauge group $SU(3)$.
    }
\end{table}
The analysis may be easily generalized to an
arbitrary gauge group with $\nf$ Dirac fermions in any
given representation.
Once again, the numerical results are
approximately reproduced by a relatively simple analytic form
which is the result of a one term variational calculation,
\begin {equation}
    \eta \simeq
    270 \, \da \, \zeta(5)^2 
    \left({2 \over \pi}\right)^5
    (v^\trans c^{-1} v) \,
    {T^3 \over g^4 \, \ln g^{-1}} \,,
\la {eq:etaapprox}
\end {equation}
where $c$ is the $2 \times 2$ matrix
\begin {equation}
   c = (\da\ca+\nf\df\cf)
       \left[ \begin {array}{cc}
	   \da\ca & 0 \\ 0 & {7\over4}\nf\df\cf
	\end {array} \right]
     + {9\pi^2\over 128} \,\nf\df\,\cf^2\,\da
       \left[ \begin {array}{rr} 1 & -1 \\ -1 & 1 \end{array} \right] ,
\la{eq:c}
\end {equation}
and
\begin {equation}
   v = \left[ \begin {array}{c} \da \\ {15\over8} \nf\df \end{array} \right] .
\la{eq:v}
\end {equation}
Here, $\df$ and $\cf$ denote the dimension and quadratic Casimir
of the fermion representation, while $\da$ and $\ca$ are the dimension
and Casimir of the adjoint representation.
(See Appendix \ref{app:group}.)
In all cases studied,
the expression (\ref {eq:etaapprox}) is accurate to within 0.7\%.
The two-by-two matrix structure of expression (\ref {eq:etaapprox})
arises from the fact that the leading-log shear viscosity is sensitive
to all two-particle scattering processes:
fermion-fermion, fermion-gluon, and gluon-gluon.
In particular, the non-diagonal second term in Eq.~(\ref{eq:c}) arises
from Compton scattering and $q \overline{q}$ annihilation to gluons,
as will be described in sections \ref{sec:leading-log} and
\ref{sec:shear}.

An earlier result of
Baym, Monien, Pethick, and Ravenhall~\cite{BMPRa}
coincides with Eqs.~(\ref{eq:etaapprox})--(\ref{eq:v})
except that they omitted this second term in the matrix $c$.
A later paper of Heiselberg \cite{heiselberg}
essentially agrees with our leading-log result for the shear viscosity
of pure gauge theory,%
\footnote
    {%
    The sign of the difference between the one-term \ansatz\ result
    and the correct leading-log coefficient is reported incorrectly
    in Appendix A of Ref.~\cite {heiselberg}.
    In a related matter, Ref.\ \cite{heiselberg} incorrectly asserts that
    the exact $\eta$ is a minimum of the variational problem set up in that
    paper; it is actually a maximum.
    }
but the later treatment of fermions in this paper also 
missed the Compton scattering and annihilation contributions, and made
additional errors not present in \cite{BMPRa}.

Plugging $\ca = 0$ and $\da = \df = \cf = 1$
into Eq.~(\ref {eq:etaapprox})
yields (a good approximation to) the leading-log QED result
for $\nf$ charged leptons and no quarks,
which is again accurate to within 0.7\%.
For an $e^+e^-$ plasma ($\nf = 1$),
this gives
\begin {equation}
    \eta 
    \simeq 270 \, \zeta(5)^2 
    \left({2 \over \pi}\right)^5
    \left(
	{529 \over 112}
	+
	{128 \over 9\pi^2}
    \right)
    {T^3 \over e^4 \, \ln e^{-1}}
    =
    187.129 \, {T^3 \over e^4 \, \ln e^{-1}}
    \, .
\la{eq:visc_QED}
\end {equation}
The complete leading-log calculation for this case gives
$\eta = 188.38 \, T^3 / (e^4 \, \ln e^{-1})$.

When applied to the standard model
at temperatures small compared to $M_W$
(but large compared to $\Lambda_{\rm QCD}$),
the results of Table \ref {shear_table} or Eq.~(\ref {eq:etaapprox}),
with $g$ the running QCD coupling (evaluated at a scale of order $T$),
are relevant for hydrodynamic fluctuations in a quark-gluon plasma
occurring on length scales which are large compared to the strong interaction
transport mean free path
of order $(g^4 T \ln g^{-1})^{-1}$,
but small compared to the electromagnetic 
transport mean free path of order $(e^4 T \ln e^{-1})^{-1}$.
In this regime, leptons may be regarded as freely streaming and
decoupled from the quark-gluon plasma.

On longer length scales, large compared to the electromagnetic 
transport mean free path, the shear viscosity is dominated by
electromagnetic scatterings of out-of-equilibrium charged leptons.
[Photons do not contribute significantly because they are thermalized by 
$\gamma g\rightarrow q\overline{q}$ and 
$\gamma q \rightarrow qg$ processes,
whose rates are $O(\alphaEM \, \alphas)$ and hence
rapid compared to purely electromagnetic scatterings.]
In this domain, the shear viscosity is approximately given by
\begin {equation}
    \eta \simeq
    (5/2)^3 \, \zeta(5)^2
    \left({12 \over \pi} \right)^5
    \left({N_{\rm leptons} \over 9\pi^2 + 224 N_{\rm species}}\right)
    {T^3 \over e^4 \, \ln e^{-1}} \, .
\la {eq:etaEMapprox}
\end {equation}
This form reproduces the correct leading-log coefficient to within
0.5\%, but neglects strong interaction effects which give
relative corrections suppressed by $\alpha_s$ or
$\alpha_{\rm EM} / \alphas$ 
(as well as next-to-leading log corrections
formally down by $1/\ln \alphaEM^{-1}$).
Once again, $N_{\rm leptons}$ is the number of light leptonic species,
and $N_{\rm species}$ is the sum over all light fermion fields weighted
by the square of their electric charges.

The result (\ref {eq:etaEMapprox}) assumes that neutrinos may
still be regarded as freely streaming and decoupled,
so it is valid only on length scales small compared to the neutrino
mean free path, which is of order $(M_Z/T)^4 /(\alpha_W^2 T)$.
On scales large compared to the neutrino mean free path, the shear
viscosity is dominated by neutrino transport and scales as $T^4$
times this mean free path,
\begin {equation}
    \eta = O\!\left( {M_Z^4 \over \alpha_W^2 T} \right) .
\end {equation}
We have not calculated the precise coefficient and, to our knowledge,
no quantitative calculation of neutrino viscosity is available
in the literature.%
\footnote
    {%
    A fairly careful estimate of the neutrino mean free path
    has been given in Ref.~\cite {HecklerHogan}.
    }

Finally, in the high temperature electroweak phase
the viscosity, like the hypercharge conductivity,
is dominated by right handed lepton transport.
Neglecting relative order $\tan^4 \theta_W$ and $\alpha'/\alpha_s$
corrections plus negligible Yukawa coupling effects,
the leading-log shear viscosity in this regime
is given by Eq.~(\ref{eq:etaEMapprox}) with $e$ replaced by $g'$,
$N_{\rm leptons} \rightarrow 3/2$,
and $N_{\rm species} \rightarrow 5+n_{\rm s}/2$.

\subsubsection {Baryon and lepton number diffusion}

To leading-log order, the fermion number diffusion constant
in a QED or QCD-like gauge theory has the form
\begin {equation}
    D_{\rm F} = \diffC_{\rm F} \, { T^{-1} \over g^4 \, \ln g^{-1} } \,,
\end {equation}
where $g$ is the gauge coupling and $\diffC_{\rm F}$ is a constant.
For the case of $SU(3)$ gauge theory,
our results for the fermion number diffusion coefficient $\diffC_{\rm F}$
for various numbers of fermion flavors
are shown in table~\ref {diff_table}.

\begin{table}[t]
\tabcolsep 10pt
\begin {center}
\begin{tabular}{cc}
$ \quad \nf \quad $ & $ D_{\rm F} \times (g^4 T) \ln g^{-1} $ \\ \hline 
``0'' & 16.0597  \\
1 & 14.3677  \\
2 & 12.9990  \\
3 & 11.8688  \\
4 & 10.9197  \\
5 & 10.1113  \\
6 & \phantom{1}9.4145
\end{tabular}
\end {center}
\caption
    {%
    \label{diff_table}
    Leading-log diffusion constant for fermion number density
    as a function of the number of fermion flavors with $m \ll T$
    in $SU(3)$ gauge theory.
    The line labeled ``0'' shows the result when only scattering
    off thermal gluons is included.
    }
\end{table}

As will be discussed in detail below,
the diffusion constant depends on the rate at which
a fermion scatters off either another fermion,
or a gluon (photon) present in the high temperature plasma.
In Table \ref {diff_table}, the line labeled ``0'' is analogous
to a quenched (or valence) approximation, and shows the result
when only scattering off thermal gluons is included.
The fairly weak dependence of the coefficient on $\nf$ shows that
in all cases the gluonic scattering contribution is dominant.

As in the previous cases, these results for an $SU(3)$ gauge theory are
approximately reproduced by the simple analytic form
\begin {equation}
    D_{\rm F} \simeq
    \left({ 2^4 \, 3^6 \, \zeta(3)^2 \, \pi^{-3} 
	\over 24 + 4 \nf + \pi^2 }
    \right)
    {T^{-1} \over g^4 \, \ln g^{-1}} \,,
\la {eq:D_F}
\end {equation}
which is the result of a one-term variational approximation.
In all cases studied this expression is accurate to within 0.3\%.
The three factors in the denominator arise, in order, from $t$-channel
gluon exchange with a gluon, $t$-channel gluon exchange with
a quark, and Compton scattering or annihilation to gluons.

This expression can be generalized to arbitrary simple gauge group and
matter fields in any representation.
The leading-log diffusion constant for the net number density
of fermion flavor $a$ is (approximately) given by
\begin{equation}
  D_a \simeq {6^5 \, \zeta(3)^2\over \pi^{3} \, C_{{\rm R}_a}} 
    \left[
	\sum_b^\ffh T_{{\rm R}_b} \lambda_b +
	{3\pi^2\over 8} \, C_{{\rm R}_a}
    \right]^{-1}
    \left( {T^{-1} \over g^4 \ln g^{-1}} \right) ,
\la{eq:approx_D}
\end{equation}
where the sum is over all flavors and helicities $b$ of the excitations
that fermion $a$ can scatter from in $ab \to ab$ processes
mediated by gauge-boson exchange,
including separately particles and antiparticles. 
(So 2 terms appear for scattering off of a gauge boson,
complex scalar, or Weyl fermion, and 4 terms for scattering
from a Dirac fermion.)
We have introduced the notation ``$\ffh$'' over the sum as a reminder
that the sum includes flavors [$\rm f$], anti-flavors if distinct
[$\rm \bar f$], and helicities [$\rm h$].
The group representation normalization factor
$T_{\rm R} \equiv C_{\rm R} d_{\rm R}/\da$
is defined in appendix \ref{app:group}.
The symbol $\lambda_b$ is
\begin {equation}
   \lambda_b = \cases{2, & if $b$ is a boson; \cr
                      1, & if $b$ is a fermion.}
\end {equation}
If the relevant scattering is by photon exchange,
then $\da = d_R = 1$, and $e_a^2 \equiv g^2 C_{{\rm R}_a}$ is the
squared electric charge of species $a$.
The result (\ref {eq:approx_D}) neglects any Yukawa interactions with
scalar fields.
Once again, the sum appearing in Eq.~(\ref {eq:approx_D}) also
appears in the lowest-order expression for the high temperature Debye mass,
now generalized to an arbitrary simple gauge group
and arbitrary matter content,
\begin {equation}
    m_D^2
    =
    {1 \over 12}
    \left( \sum^\ffh_b T_{{\rm R}_b} \lambda_b \right) g^2 T^2 \,.
\end {equation}

When applied to the standard model
at temperatures small compared to $M_W$,
the result (\ref {eq:D_F}) with $g^2 = \gs^2$
gives (a good approximation to) the leading-log result for the diffusion 
constant which is appropriate for describing relaxation
of fluctuations in baryon density, or equivalently the
net density of any particular quark flavor, on scales which
are large compared to the
strong interaction transport mean free path
of order $(\gs^4 T \ln \gs^{-1})^{-1}$.
If the departure from equilibrium of the various quark densities
has vanishing electric charge density, then the resulting relaxation
is purely diffusive.
However, if the perturbation in quark densities
has a net non-zero electric charge density,
then electromagnetic interactions can only be neglected if the
scale of the fluctuation is small compared to the the
electromagnetic transport mean free path
of order $(e^4 T \ln e^{-1})^{-1}$.
On longer scales, the net charge density will relax at a rate
affected by the electrical conductivity, while electrically
neutral flavor asymmetries will relax diffusively.
This will be discussed further momentarily.

The leading-log diffusion constant characterizing
the relaxation of fluctuations in charged lepton densities
which are electrically neutral
[{\em e.g.}, an excess of electron minus positron density,
balanced by an equal excess in $\mu^+ - \mu^-$ number density]
is given by the appropriate specialization of
Eq.~(\ref {eq:approx_D}) to QED, namely
\begin {equation}
    D_{L}
    \simeq
    {6^5 \, \zeta(3)^2\, \pi^{-3} } 
    \left[
	0 + 4 N_{\rm species} + \frac{3 \pi^2}{8}
    \right]^{-1}
    \left( {T^{-1} \over e^4 \ln e^{-1}} \right) \,.
\la {eq:D_L}
\end {equation}
Once again, $N_{\rm species}$ is the sum over all relevant fermion
fields weighted by the square of their electric charge.
In Eq.~(\ref {eq:D_L}), and in the following
Eqs.~(\ref {eq:D_LL})--(\ref {eq:D_B2}),
the first term in the square bracket arises from $t$ channel scattering
from a gauge boson, the middle term represents $t$ channel scattering from 
something else, and the third term arises from Compton scattering and 
annihilation to gauge bosons.

The relaxation of an arbitrary set of slowly-varying fluctuations $n_a$
in net quark and (charged) lepton densities
(with $a$ labeling both quark and lepton species)
is described by the coupled set of diffusion/relaxation equations
\begin {equation}
    {\partial n_a \over \partial t} = D_a \, \nabla^2 n_a
    - {\sigma_a \over e_a} \sum_b e_b \, n_b \,,
\la {eq:charged-diffusion}
\end {equation}
where $\sigma_a$ is the contribution to the conductivity due to
charge carriers of species $a$
(so that $e_a \,\j_a = \sigma_a \E$, where
$\j_a$ is the species $a$ particle number flux,
and the total conductivity $\sigma = \sum_a \sigma_a$).
These equations encode the fact that, in addition to
various diffusive processes,
the charge density $\rho \equiv \sum_a e_a n_a$
satisfies the non-diffusive relaxation equation $\dot \rho = -\sigma \rho$,
up to second order gradient corrections, showing that
the conductivity is the relaxation rate for large-scale
charge density fluctuations.%
\footnote
    {%
    This, of course, immediately follows from combining the
    continuity equation for electric charge
    $\dot \rho + \nabla \cdot \j = 0$,
    the defining relation for conductivity $\j = \sigma \E$,
    and Gauss' law $\nabla \cdot \E = \rho$.%
    }
In fact
(as noted by Einstein),
the conductivity is directly related to the
underlying diffusion constants of individual species
through the simple relation%
\footnote{
  This identity may be seen directly from the Kubo relations
  (\ref {eq:sigmaKubo}) and (\ref {eq:DKubo}).
  Alternatively,
  a simple physical derivation
  (specializing for convenience to electromagnetism with a single species)
  is easily given.
  Start with the diffusion equation
  $\j = - e D \grad n = - e D (dn/d\mu) \grad\mu$.
  Then realize that in a constant electric field
  the effective chemical potential is
  $\mu = \mu_0 - e \E\cdot\x$.
  Hence $\j = e^2 D (dn/d\mu) \E$, or
  $\sigma = e^2 D (dn/d\mu)$.
}
\begin {equation}
    \sigma = \sum_a e_a^2 \, D_a \, {\partial n_a \over \partial \mu_a}
\la {eq:einstein}
\end {equation}
[or $\sigma_a = e_a^2 \, D_a \, (\partial n_a / \partial \mu_a)$].
This relation assumes that the the charge susceptibility
matrix (\ref {eq:susceptability}) is diagonal,
as it is when all charge densities vanish.
For (effectively) massless fermions,
$\partial n/\partial\mu = {1\over3} \, T^2$ per
Dirac fermion.  Since $D_L \gg D_F$, leptons completely dominate over
quarks in the above species sum.
Inserting the result (\ref {eq:D_L}) for the lepton diffusion
constant into the Einstein relation (\ref {eq:einstein})
reproduces our previous expression (\ref {eq:approx-cond}) for
conductivity, as it must.

For temperatures large compared to $M_W$
({\em i.e.}, in the high temperature electroweak phase),
one may find corresponding results for lepton number diffusion
by specializing the general result (\ref {eq:approx_D}).
If Yukawa interactions are neglected,
then the relaxation of left and right-handed lepton number
excesses are independent.
The diffusion constant for left-handed net lepton number
in high temperature electroweak theory is controlled by the
$SU(2)_L$ gauge interactions and (approximately) equals
\begin {equation}
    D_{L_L}
    \simeq
    {6^5 \, \zeta(3)^2\, \pi^{-3} } 
    \left[
	6 + \frac{3}{4} \, (\nf +2 n_{\rm s}) + \frac{27 \pi^2}{128}
    \right]^{-1}
    \left( {T^{-1} \over g_w^4 \ln g_w^{-1}} \right) ,
\la {eq:D_LL}
\end {equation}
with relative corrections of order $1/\ln \gw^{-1}$,
$\tan^2 \theta_W = (g'/\gw)^2$, and $\alphas$.
Here, $\nf=12$ denotes the number of $SU(2)$ chiral doublets,
and $n_s$ is the number of scalar doublets.
The corresponding diffusion constant for right-handed lepton number
depends on hypercharge interactions,
\begin {equation}
    D_{L_R}
    \simeq
    {6^5 \, \zeta(3)^2\, \pi^{-3} } 
    \left[
	0 + (20 + 2 n_{\rm s}) + \frac{3 \pi^2}{8}
    \right]^{-1}
    \left( {T^{-1} \over g'{}^4 \ln g'{}^{-1}} \right) ,
\la {eq:D_LR}
\end {equation}
with relative corrections of order $1/\ln g'{}^{-1}$ and $\alphas$.
Inclusion of Yukawa interactions will cause the diffusion of
right and left-handed lepton number excesses to become coupled.
This, however, only becomes relevant on scales larger than
the mean free path for scattering processes involving
Higgs emission, absorption or exchange.
In the minimal standard model, this scale is of order
$[(m_\ell/M_W)^2 (m_{\rm t}/M_W)^2 \alphaw^2 T]^{-1}$,
where $m_\ell$ is the (zero temperature) mass of the lepton species
of interest,
and $m_{\rm t}$ is the top quark mass.
In other words, this scale is larger than the $SU(2)_L$ transport
mean free path by a factor of roughly $(M_W/m_\ell)^2$.

The baryon diffusion constant for $T \gg M_W$
is still given by the previous result (\ref {eq:D_F}) [with $g = \gs$],
up to relative corrections of order $1/\ln \gs^{-1}$.
This may be rewritten as
\begin {equation}
    D_B
    \simeq
    {6^5 \, \zeta(3)^2\, \pi^{-3} } 
    \left[
	16 + {16\over 3} \, N_g + \frac{2 \pi^2}{3}
    \right]^{-1}
    \left( {T^{-1} \over \gs^4 \ln \gs^{-1}} \right) ,
\la {eq:D_B2}
\end {equation}
where $N_g = 3$ is the number of generations.

These (approximations to leading-log) diffusion constants for
baryon and left or right-handed lepton number density characterize
the relaxation in the high temperature electroweak phase of
arbitrary fluctuations in any of these densities which
are hypercharge neutral.
For fluctuations having non-zero hypercharge density,
one must also include the effect of the induced
hypercharge electric field,
leading to the same coupled relaxation equations as in
(\ref {eq:charged-diffusion}),
but with $\sigma$ now the hypercharge conductivity
(and the species indices $a,b$ now labeling right-handed leptons,
left-handed charged leptons, and quark flavors).
For the hypercharge conductivity, the dominant contribution
to the Einstein relation (\ref {eq:einstein}) comes from
right-handed leptons (since $g'$ is the weakest coupling).
Once again, one may easily check that inserting the result (\ref {eq:D_LR})
for the right-handed lepton diffusion constant into the Einstein
relation reproduces our previous expression (\ref {eq:hypercond})
for the hypercharge conductivity.

We have not computed diffusion
constants for conserved numbers carried by scalars.

Several determinations of leading-log diffusion
constants have previously been reported 
\cite{Heiselberg_diff,JPT1,MooreProkopec,JPT2}.  Of these, only 
Moore and Prokopec \cite{MooreProkopec} included all relevant
diagrams, and each of these previous calculations made errors in
evaluating at least one diagram.
We will discuss the differences between our treatment and these
previous results in more detail in Sec.~\ref{sec:diffusion}.

\subsubsection {$U(N_f) \times U(N_f)$ flavor diffusion}

In QCD (or any QCD-like theory) with $\nf$ species of fermions,
one may consider the entire set of
$SU(\nf)_\V \times SU(\nf)_\A \times U(1)_\B \times U(1)_\A$
currents.
The diagonal components of the $SU(\nf)_\V$ currents are exactly
conserved (neglecting weak interactions), while
in our high temperature regime, the off-diagonal $SU(\nf)_\V$
currents are approximately conserved if one neglects order $(\Delta m/T)^2$
effects, where $\Delta m$ is some fermion mass difference.
Similarly, conservation of the $SU(\nf)_\A$ currents is
spoiled only by $O(m^2/T^2)$ corrections.
Consequently,
the constitutive relations for the different currents must decouple,
\begin {eqnarray}
    \hbox {$U(1)_\B$} \; &:& \qquad
    \langle \j_\B \rangle = -D_\B \, \nabla \langle n_\B \rangle \,,
\\
    \hbox {$SU(\nf)_\V$} \; &:& \qquad
    \langle \j_\V^\alpha \rangle = -D_\V \,\nabla \langle n_\V^\alpha \rangle\,,
\\
    \hbox {$SU(\nf)_\A$} \; &:& \qquad
    \langle \j_\A^\alpha \rangle = -D_\A \,\nabla \langle n_\A^\alpha \rangle\,,
\\
    \hbox {$U(1)_\A$} \; &:& \qquad
    \langle \j_\A \rangle = -D'_\A \, \nabla \langle n_\A \rangle \,,
\la {eq:U(1)_A diffusion}
\end {eqnarray}
and the full diffusion constant matrix (in this basis) is diagonal
when power corrections vanishing like $T^{-2}$,
as well as weak and electromagnetic interactions, are neglected.

The $U(1)_\B$ baryon number current is almost
exactly conserved,%
\footnote
    {%
    Baryon number is exactly conserved in QCD, but its conservation
    is violated by electroweak effects, at rates that are exponentially
    small at $T \lsim M_W$ \pcite{tHooft,ArnoldMcLerran} and
    $O(\alpha_w^5 T \ln \alpha_w^{-1})$ at
    very high temperatures \pcite{ASY0,bodeker,ASY}.%
    }
so fluctuations in baryon number density will behave diffusively on 
length and time scales large compared to the appropriate mean free
scattering time.
Fluctuations in flavor asymmetries ---
that is, the diagonal components of the $SU(\nf)_\V$ current densities ---
will behave diffusively on time scales large compared to QCD
mean free scattering times but small compared to the mean free
time for flavor-changing weak interactions, which is of order
$(m_W/T)^4 / (\alpha_w^2 T)$,
or (if the fluctuation is electrically charged) the electromagnetic
transport mean free time of order $(e^4 T \ln e^{-1})^{-1}$.

Fluctuations in the $SU(\nf)_\A$ and off-diagonal $SU(\nf)_\V$
charge densities will behave diffusively on time scales large compared
to mean free transport scattering times but small compared to the
time scale of order $T/m^2$ or $T/(\delta m)^2$
where the respective symmetry breaking interactions become relevant.%
\footnote
    {%
    So in the presence of non-zero fermion masses, or mass differences,
    at sufficiently high temperature fluctuations in the
    $SU(\nf)_\V$ and $SU(\nf)_\A$ charge densities are
    ``pseudo''-diffusive modes, analogous to pseudo-Goldstone bosons.
    Large scale fluctuations in these ``almost-conserved'' charge densities
    will satisfy a diffusion/relaxation equation of the form
    $\dot n = D \nabla^2 n - \gamma n$,
    where the local relaxation rate $\gamma$ will be
    $O[(\delta m)^2/T]$ or $O(m^2/T)$, respectively.%
    }
Because of the axial anomaly, fluctuations in the $U(1)_\A$
axial charge density may relax locally on a time scale of order
$(\alpha_s^5 T \ln \alpha_s^{-1})^{-1}$
even in the massless theory.
The physics behind this is completely analogous to treatment
of baryon violating transitions in high temperature electroweak theory
\cite {ASY0,bodeker,ASY,top-trans-other1,top-trans-other2}.%
\footnote
    {%
    A useful discussion of this material may be found in Ref.~\cite {MMS}
    [which, however, predates the realization that the transition
    rate per unit volume scales as $O(\alpha^5 T^4 \ln \alpha^{-1})$,
    not as $(\alpha T)^4$].%
    }
As for the other axial currents, fluctuations in $U(1)$ axial charge
density will relax diffusively on time scales large compared to the
QCD transport mean free times, but small compared to both
the perturbative $T/m^2$ and non-perturbative
$(\alpha_s^5 \, T \ln \alpha_s^{-1})^{-1}$
scales where $U(1)_\A$ violation becomes apparent.
Axial charge fluctuations within this domain may be characterized by
the basic diffusion equation (\ref {eq:U(1)_A diffusion}), with
a diffusion constant $D_\A'$ which is perturbatively computable.

As will be discussed in detail in section \ref {sec:diffusion},
to leading order in $\alpha_s$ the various flavor diffusion constants
only depend on two-to-two particle scattering rates
in the high temperature plasma.
Consequently, the diffusion constants for currents corresponding
the various (approximate) flavor symmetry groups are all identical,
\begin {equation}
    D_\B = D_\V = D_\A = D'_\A \,,
\end {equation}
up to relative corrections suppressed by one or more powers of $\alphas$.

\section {kinetic theory and transport coefficients}
\la {sec:kinetic}

    To calculate any of the transport coefficients under consideration,
correct to leading order in the interaction strength $g$
but valid to all orders in $1/\ln g^{-1}$, it is sufficient to use a kinetic
theory description for the relevant degrees of freedom.
One introduces a particle distribution function $f(\p,\x,t)$
characterizing the phase space density of particles
(which one should think of as coarse-grained on a scale
large compared to $1/T$, but small compared to mean free paths).
The distribution function $f(\p,\x,t)$ is really a multi-component
vector with one component for each relevant particle species
(quark, gluon, {\em etc}.), but this will not be indicated explicitly
until it becomes necessary.
The distribution function satisfies a Boltzmann equation of the usual form
\begin {equation}
    \left[
	{\partial \over \partial t}
	+
	\v_\p \cdot {\partial \over \partial \x}
	+
	{\bf F_{\rm ext}} \cdot {\partial \over \partial \p}
    \right]
    f(\p,\x,t)
    =
    -C[f] \,.
\la {eq:Boltz}
\end {equation}
The external force ${\bf F}_{\rm ext}$ term will only be relevant
in discussing the electrical conductivity.
Since typical excitations in the plasma [those with $\p = O(T)$]
are highly relativistic, corrections to their dispersion relations
are suppressed by $O(g^2)$, and may be neglected.
Consequently, one may treat all excitations as moving at the speed
of light, which means that the spatial velocity is a unit vector,
$\v_\p = \hat \p \equiv \p / |\p|$.%
\footnote
    {%
    This assumes, of course, that the particular physical quantities
    under consideration are dominantly sensitive to the behavior of
    typical ``hard'' excitations.
    We will see that this is true for the transport coefficients under
    discussion.  
    However, certain other observables, such as the bulk viscosity,
    may be sufficiently sensitive to the dynamics of ``soft'' excitations
    with momenta $\p \ll T$ that their calculation requires
    an improved treatment which adequately describes both
    hard and soft degrees of freedom.%
    }

For calculations to leading order in $g$, and for the transport
coefficients under consideration, it will be sufficient to
include in the collision term $C[f]$ only two-body scattering processes,
so that
\begin {eqnarray}
    C[f](\p)
    &=&
    \half
    \int_{\k,\p',\k'}
	\left| M(p,k,p',k') \right|^2 \>
	(2\pi)^4 \, \delta^4 (p+k-p'-k')
\nonumber\\ && \quad\quad {} \times
    \Bigl\{
	f(\p) \, f(\k) \, [1{\pm}f(\p')] \, [1{\pm}f(\k')]
	-
	f(\p') \, f(\k') \, [1{\pm}f(\p)] \, [1{\pm}f(\k)]
    \Bigr\} \,.
\la {eq:collision}
\end {eqnarray}
Here, $p$, $k$, {\em etc}., denote on-shell four-vectors
(so that $p^0 = |\p|$, {\em etc.}),
$M(p,k,p',k')$ is the two body scattering amplitude
with non-relativistic normalization,
related to the usual relativistic amplitude $\cal M$ by
\begin {equation}
    |M(p,k,p',k')|^2 = {|{\cal M}(p,k,p',k')|^2 
	\over (2p_0)(2k_0)(2p'_0)(2k'_0)} \,,
\end {equation}
and $\int_\p$ is shorthand for $\int d^3 \p / (2\pi)^3$.
The collision term is local in spacetime, and all distribution functions
are to be evaluated at the same spacetime position (whose coordinates
have been suppressed).
With multiple species of excitations there will, of course,
be species-specific scattering amplitudes
and multiple sums over species.
As always, in the $1{\pm}f$ final state statistical factors,
the upper sign applies to bosons and the lower to fermions.
Appropriate approximations for the scattering amplitudes will be discussed
in the next section.

The stress-energy tensor, in this kinetic theory, equals
\begin {equation}
    T^{\mu\nu}(x) = \int_\p v_\p^\mu \> p^\nu \, f(\p,x) ,
\la {eq:kineticT}
\end {equation}
where $v_\p^\mu \equiv p^\mu / p^0$ is a convenient generalization of the
three-vector velocity for
an excitation with spatial momentum $\p$.
($v_\p^\mu$ is not the four-velocity and transforms non-covariantly,
in just the manner required so that
$(d^3\p) \, v_\p^\mu$ does transform covariantly.)
Other conserved currents
are given by similar integrals over the
phase-space distribution function, but with the implied species sum
weighted by appropriate charges $q$ of each species,%
\footnote
    {%
    Eq.\ (\ref{eq:kineticJ}) is adequate for currents which are diagonal
    in the basis of species.
    More generally, the distribution function $f(\p,x)$ should be
    viewed as a quantum density matrix for all internal degrees of freedom
    of an excitation.
    For example, in a theory with particles transforming in some
    representation $R$ of the global symmetry group, the distribution
    function transforms under the $R \times \bar R$ representation.
    Off-diagonal components of the distribution function are
    relevant if one is interested in, for example, the off-diagonal
    parts of the $SU(\nf) \times SU(\nf)$ currents.
    Eq.\ (\ref{eq:kineticJ}) would then be generalized to
    $ j_\alpha^\mu(x) = \int_\p v_\p^\mu \;
      {\rm tr} \left[ t^\alpha f(\p,x) \right]$,
    where $t^\alpha$ is the appropriate charge representation matrix.
    }
\begin {equation}
    j^\mu(x) = \int_\p v_\p^\mu \;
    q \, f(\p,x) .
\la {eq:kineticJ}
\end {equation}
The factor $p^\nu$ which appears with $f(\p,x)$ in Eq.~(\ref{eq:kineticT}) 
reflects the fact that in this case it is the energy or momentum of
an excitation which is the conserved charge.
Given the Boltzmann equation (\ref {eq:Boltz}), and 
scattering amplitudes in (\ref {eq:collision})
which respect the microscopic conservation laws,
one may easily check that the currents
(\ref {eq:kineticT}) and (\ref {eq:kineticJ}) are, in fact, conserved.

To extract transport coefficients, it is sufficient to linearize
the Boltzmann equation (\ref {eq:Boltz}) and examine the response
of infinitesimal fluctuations in various symmetry channels.
This will be described explicitly below.

But first we digress to discuss the validity of this kinetic theory approach.
The Boltzmann equation (\ref {eq:Boltz})
may be regarded as an effective theory,
produced by integrating out (off-shell) quantum
fluctuations, which is appropriate for describing the dynamics
of excitations on scales large compared to $1/T$, which is the size of the
typical de Broglie wavelength of an excitation.
The use of kinetic theory for calculating transport coefficients
may be justified in at least three different ways:
\begin {enumerate}
\item
     One may begin with the full hierarchy of Schwinger-Dyson equations for
     (gauge-invariant) correlation functions in a weakly non-equilibrium state
     in the underlying quantum field theory.
     For weak coupling, one may systematically justify, and then insert,
     a quasi-particle approximation for the spectral densities of the
     basic propagators, perform a suitable gradient expansion and
     Wigner transform,
     and formally derive the above kinetic theory.
     (See Refs.~\cite {Blaizot&Iancu,kinetic-thy1,kinetic-thy2}
     and references therein.)
\item
    One may consider the diagrammatic expansion for the equilibrium
    correlator appearing in the Kubo relation (\ref {eq:Kubo})
    for some particular transport coefficient.
    After carefully analyzing the contribution of arbitrary diagrams
    in the kinematic limit of interest ($\k=0$ and $\omega \to 0$),
    one may identify, and resum, the infinite series of diagrams
    which contribute to the leading-order result.
    One obtains a linear integral equation, which will coincide exactly
    with the result obtained from linearizing the appropriate kinetic theory.
    This program has been carried out explicitly for scalar theories
    \cite{JeonYaffe,Jeon} but not yet for gauge theories.
\item
    One may directly argue (by examining equilibrium finite temperature
    correlators) that, for sufficiently weak coupling,
    the underlying high temperature quantum field theory
    has well-defined quasi-particles,
    that these quasi-particles are weakly interacting
    with a mean free time large compared to the actual
    duration of an individual collision, and consequently
    that scattering amplitudes of these quasi-particles are well-defined
    to within a precision of order of the ratio of these scales.
    In other words, one justifies the existence
    of quasi-particles by looking at the spectral densities
    of the propagators of the basic fields, reads off their
    scattering amplitudes from looking at higher point correlators,
    and writes down the kinetic theory which correctly describes
    the resulting quasi-particle interactions.
    For a more detailed discussion of this approach see, for example,
    Ref.~\cite {A&Y}.
\end {enumerate}
Given the complexities of real-time, finite-temperature diagrammatic
analysis in gauge theories (especially non-Abelian theories),
we find the last approach to be the most physically transparent and compelling.
But this is clearly a matter of taste.

There is one important caveat in the claim that a kinetic theory
of the form (\ref {eq:Boltz}) can accurately describe excitations
in a hot gauge theory.
One may argue, as just sketched, that such a Boltzmann equation
with massless dispersion relations reproduces
(to within errors suppressed by powers of $g$)
the dynamics of {\em typical} excitations in the plasma,
namely hard excitations whose momenta are of order $T$.
For such excitations, thermal corrections to the massless lowest-order
dispersion relations are a negligible $O(g^2)$ effect.
This is not true for soft excitations with momenta of order $gT$, or less.
In gauge theories,
one cannot characterize sufficiently long wavelength dynamics
in terms of (quasi)-particle excitations with purely local collisions.
Instead, one may think of long wavelength degrees of freedom
as classical gauge field fluctuations, and construct Boltzmann-Vlasov
type effective theories which describe hard excitations propagating
in a slowly varying classical background field.
The well known hard-thermal-loop (HTL) effective theory is of precisely
this form \cite {Heinz,BraatenPisarski,HTL1,HTL2,Blaizot&Iancu}.

A simple kinetic theory of the form (\ref {eq:Boltz}),
without the complications of background gauge field fluctuations,
can only be adequate for computing physical quantities which are
not dominantly sensitive to soft excitations.
This is true of most observables, including thermodynamic quantities
such as energy density or entropy, just because phase space grows as
$\p^2 d|\p|$ in (3{+}1) dimensional theories.
This is equally true for the transport coefficients under consideration.
It will be easiest to demonstrate this {\em a-posteriori}.
However, this insensitivity (at leading order) to soft excitations
may not hold for the bulk viscosity, which is why its calculation
requires a more refined analysis.
(This is true even in a pure scalar theory \cite {JeonYaffe,Jeon}.)

Returning to the analysis of the Boltzmann equation (\ref {eq:Boltz}),
equilibrium solutions are given by
\begin {equation}
    f^a_{\rm eq}(\p)
    =
    \left\{
	\exp
	    \left[ \beta (-u_\nu p^\nu - \mu_\alpha \, q_\alpha^a) \right]
	\mp 1
    \right\}^{-1} ,
\end {equation}
where $\beta$ is the inverse temperature, $u$ is the fluid four-velocity,
and $\{ \mu_\alpha \}$ are chemical potentials corresponding to a
mutually commuting set of conserved charges.
We have now included an explicit species index $a$, and $q_\alpha^a$ is the
value of the $\alpha$'th conserved charge carried by species $a$.
(Sums over repeated charge indices should be tacitly understood.)

Using only the fact that the scattering amplitudes respect the
microscopic conservation laws, one may easily show that
the collision term exactly vanishes for any such equilibrium distribution,
$C[f_{\rm eq}] = 0$.

The distribution function corresponding to some
non-equilibrium state which describes a small departure from equilibrium
may be written as the sum of a local equilibrium distribution
plus a departure from local equilibrium.
This is conveniently written in the form
\begin {equation}
    f^a(\p,x)
    =
    f^a_0(\p,x) + f^a_0(\p,x) [1 \pm f^a_0(\p,x)] \, f^a_1(\p,x) \,.
\la {eq:f1}
\end {equation}
Here $f^a_0(\p,x)$ has the form of an equilibrium distribution function,
but with temperature, flow velocity, and chemical potentials which
may vary in spacetime,
\begin {equation}
    f^a_0(\p,x)
    =
    \left. f^a_{\rm eq}(\p) \right|_{\beta(x), u^\nu(x), \mu_\alpha(x)} \,.
\end {equation}
Writing the departure from local equilibrium as
$f_0 (1 \pm f_0) \, f_1$,
instead of just $\delta f$, simplifies the form of the resulting linearized
collision operator (\ref {eq:linearC}).
When inserted into the collision term of the Boltzmann equation,
the local equilibrium part of the distribution gives no contribution,
$C[f_0] = 0$, because the collision term is local in spacetime
and so cannot distinguish local equilibrium from genuine equilibrium.
Hence, the collision term, to first order in the departure from
equilibrium, becomes a linear operator acting on the departure
from local equilibrium,
\begin {equation}
    C[f]
    =
    {\cal C} f_1 + O(f_1^2) \,,
\end {equation}
where the action of the linearized collision operator $\cal C$ is given by
\begin {eqnarray}
    \left({\cal C} f_1\right)^a(\p)
    &\equiv&
    \half \,
    \sum_{bcd}^\ffhc \int_{\k,\p',\k'}
	\left| M^{ab}_{cd}(p,k,p',k') \right|^2 \>
	(2\pi)^4 \, \delta^4 (p+k-p'-k')
\nonumber\\ && {} \times
	f^a_0(\p) \, f^b_0(\k) \,
	[1{\pm}f^c_0(\p')] \, [1{\pm}f^d_0(\k')]
	\left[
	    f^a_1(\p) + f^b_1(\k) - f^c_1(\p') - f^d_1(\k')
	\right] .
\la {eq:linearC}
\end {eqnarray}
All distribution functions are evaluated at the same point in spacetime,
whose coordinates have been suppressed.
$M^{ab}_{cd}(p,k,p',k')$ is the scattering amplitude
for species $a$ and $b$, with momenta $\p$ and $\k$, respectively,
to scatter into species $c$ and $d$ with momenta $\p'$ and $\k'$.
For reference, this choice of momentum and species labels is
summarized in Fig.\ \ref{fig:labels}.
The ``c'' in the ``$\ffhc$'' above the sum indicates that, in the
application to gauge theories, the sum is over all
colors of the particles represented by $b$, $c$, and $d$,
as well as flavors, anti-flavors (where distinct), and helicities.

\begin{figure}[t]
\centerline{\epsfxsize=2in\epsfbox{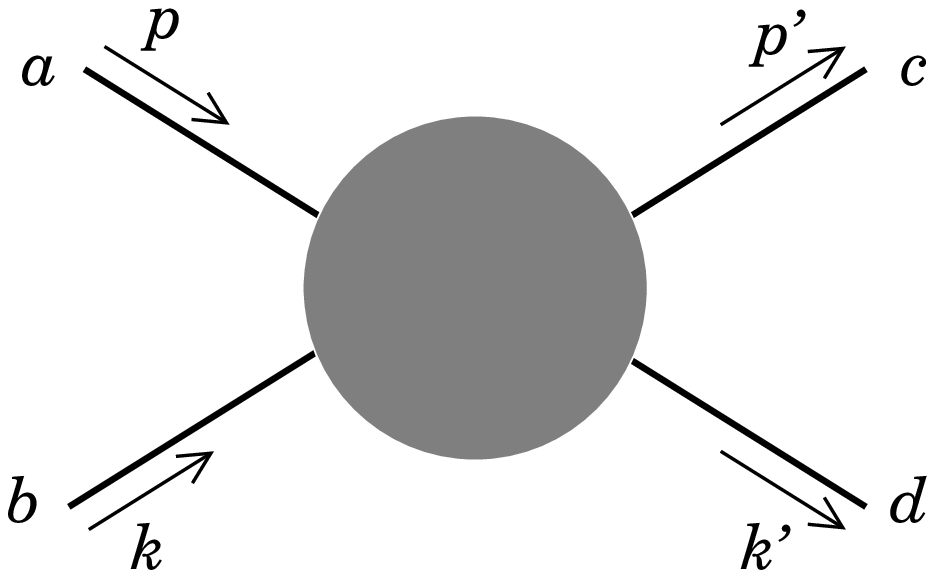}}
\vspace{0.1in}
\caption{%
    \la{fig:labels}
    Momentum and species label conventions for $2\to2$ scattering.
    Time runs from left to right.
}%
\end{figure}

On the left-hand side of the Boltzmann equation, the gradients acting on $f_0$
give a result whose size is set by either the magnitude of spacetime
gradients in temperature, velocity, or chemical potentials, or by
the magnitude of the imposed external force.
So, to first order in the departure from equilibrium,
the Boltzmann equation becomes an inhomogeneous linear integral equation
for $f_1(\p)$,
\begin {equation}
    \left[
	{\partial \over \partial t}
	+
	\hat \p \cdot {\partial \over \partial \x}
	+
	{\bf F}^a_{\rm ext} \cdot {\partial \over \partial \p}
    \right]
    f^a_0(\p,\x,t)
    =
    -\left( {\cal C} f_1 \right)^a(\p,\x,t) \,.
\la {eq:linearB}
\end {equation}
In other words, $f_1$ is first order in gradients (or the external force).
The neglected terms on the left-hand side, where derivatives act on the
deviation from local equilibrium, are of second order in gradients
(or external force), and do not contribute to the linearized analysis.

Each transport coefficient under consideration will depend on
the departure from equilibrium $f_1$ resulting from a particular
form of the driving terms on the left-hand side of the linearized
Boltzmann equation (\ref {eq:linearB}).
For conductivity, one is interested in the response to
a homogeneous electric field, where ${\bf F}^a_{\rm ext} = q^a \E$.
For diffusion or shear viscosity, one is interested in the response
to a spatial variation in a chemical potential or the fluid flow velocity.
Using the fact that
\begin {equation}
    {\rm d} f_0(\p,x)
    =
    f_0(\p,x) [1 \pm f_0(\p,x) ] \>
    {\rm d} \!\left[ \beta u_\nu p^\nu + \beta \mu_\alpha \, q_\alpha^a \right],
\end {equation}
one may easily see that in all three cases, the left hand side of the
linearized Boltzmann equation (\ref {eq:linearB}) has the form,%
\footnote
    {%
    For diffusion or viscosity,
    the time derivative term on the left-hand side may be dropped
    because its contribution is actually second order in spatial gradients.
    This follows from current (or stress-energy) conservation
    and the constitutive relations (\ref {eq:Tij}) or (\ref {eq:jia}),
    which together imply that time derivatives of the conserved
    densities are related to second spatial derivatives of the
    conserved densities themselves.
    }
\begin {equation}
    \LHS
    =
    \beta f^a_0(\p,x) [1 \pm f^a_0(\p,x) ] \>
    q^a \, I_\ij(\hat \p) \, X_\ij(x) \,,
\la {eq:LHS}
\end {equation}
where the spatial tensor $X_\ij(x)$ denotes the
``driving field,'' namely
\begin {equation}
    X_\ij(x)
    \equiv
    \cases
        {	
	-E_i \,, \vphantom {\Big|}& \mbox{(conductivity)} \cr
	\;\nabla_i \, \mu_\alpha \,, & \mbox{(diffusion)} \cr
	\frac{1}{\sqrt{6}} \left( \nabla_i u_j + \nabla_j u_i
	- {2\over3}\delta_{ij} \nabla \cdot \u \right) ,
	& \mbox{(shear viscosity)} \cr
 	}
\label {eq:Tdrive}
\end {equation}
and
$I_\ij(\hat \p)$ is the unique $\ell = 1$ or $\ell = 2$
rotationally covariant tensor depending only on the direction of $\p$,
that is
\begin {equation}
    I_\ij(\hat \p)
    \equiv
    \cases
	{
	\;\hat p_i \,,\vphantom {\Big|} & \mbox{(conductivity/diffusion)} \cr
	\sqrt {3\over2} \, (\hat p_i \hat p_j - {1\over3} \delta_{ij}) \,. 
	& \mbox{(shear viscosity)} \cr
	}
\label {eq:Iij}
\end {equation}
In Eq.~(\ref {eq:LHS}), 
$q^a$ denotes the relevant charge of species $a$ which,
in the case of shear viscosity, means the magnitude of its momentum $|\p|$.
The factor of $\sqrt{3\over2}$ included in the
definition (\ref {eq:Iij}) of $I_{ij}$
[and the corresponding $1\over\sqrt 6$ factor for $X_{ij}$
in Eq.~(\ref {eq:Tdrive})]
are inserted so that the normalization
$I_\ij I_\ij = 1$ holds for both the $\ell = 1$ and $\ell = 2$ cases.

We will henceforth always work
in the local fluid rest-frame.
At any point $x$,
the local equilibrium distribution function $f_0(\p,x)$
then depends only on the energy $p^0 = |\p|$, and thus the
only angular dependence on $\hat \p$ in (\ref {eq:LHS}) comes from the
$\ell = 1$ or $\ell = 2$ irreducible tensor $I_\ij(\hat \p)$.
Since the linearized collision operator $\cal C$ is local in spacetime and
rotationally invariant (in the local fluid rest-frame at $x$),
the departure from equilibrium must have the same
angular dependence as the driving term.
Consequently,
given a left-hand side of the form (\ref {eq:LHS}),
the function $f_1(\p,x)$ which will solve
the linearized Boltzmann equation (\ref {eq:linearB}) must have
the corresponding form
\begin {equation}
    f^a_1(\p,x)
    =
    \beta^2 \, X_\ij(x) \, I_\ij(\hat \p) \> \chi^a(|\p|) \,,
\label {eq:fa1}
\end {equation}
where, for each species $a$, $\chi^a(|\p|)$ is some rotationally
invariant function depending only on the energy of the excitation.
The factor of $\beta^2$ is inserted for later convenience,
and causes $\chi^a$ to have the same dimensions as $q^a$
(dimensionless for conductivity or diffusion, and dimension one
for viscosity).
For notational convenience we will also define
\begin {equation}
    \chi^a_\ij(\p) \equiv I_\ij(\hat \p) \> \chi^a(|\p|) \,,
\end {equation}
and
\begin {equation}
    S^a_\ij(\p) \equiv
    -T \, q^a f^a_0(\p,x) [1 \pm f^a_0(\p,x) ] \>
    I_\ij(\hat \p) \,,
\end {equation}
so that the linearized Boltzmann equation for any
particular channel under consideration can be written
in the concise form
\begin{equation}
    S^a_\ij(\p)
    =
    \left({\cal C} \chi_\ij\right)^a(\p) \,.
\la {eq:lin1}
\end{equation}

A straightforward approach for numerically solving these coupled
integral equations would be to reduce them to a set of scalar equations
[by contracting both sides with $I_\ij(\hat\p)$],
discretize the allowed values of $\magp$,
compute the matrix elements ${\cal C}^{ab}(\magp,|\q|)$ of the kernel
of the (projected) collision operator by numerical quadrature,
and thereby convert (\ref {eq:lin1})
into a finite dimensional linear matrix equation.
This is a bad strategy, however, particularly for gauge theories.
The problem is that the kernel ${\cal C}^{ab}(|\p|,|\q|)$
has (integrable) singularities and is not smooth
as $|\p|$ crosses $|\q|$.
Consequently it is quite difficult to avoid large discretization errors
and obtain good convergence to the correct answer.

A better strategy, which is applicable to the full leading-order analysis
and not just the present leading-log treatment,
is to convert the linear integral equations
(\ref {eq:lin1}) into an equivalent variational problem.
This permits one to obtain quite accurate results using
very modest basis sets.
This conversion is trivial once one notes that the linear operator
${\cal C}$ is Hermitian with respect to the natural inner product
\begin {equation}
    \Big( f,g \Big) \equiv \beta^3 \sum^\ffhc_a \int_\p \, f^a(\p) \, g^a(\p)
    \,.
\end {equation}
Consequently, if one defines the functional
\begin {equation}
    Q[\chi]
    \equiv
    \Big( \chi_\ij, S_\ij\Big)
    - \half \, \Big( \chi_\ij, {\cal C} \chi_\ij \Big) ,
\la {eq:Q1}
\end {equation}
then the maximum value of $Q[\chi]$ occurs when $\chi^a(\magp)$ satisfies
the desired linear equation (\ref {eq:lin1}).
For later use, note that the maximal value of
the functional $Q$ may be written in either of the forms
\begin {equation}
    Q_{\rm max}
    =
    \left.
	\half \, \Big( \chi_\ij, {\cal C} \chi_\ij \Big) 
    \right|_{\chi = \chi_{\rm max}}
    =
    \left.
	\half \, \Big(\chi_\ij, S_\ij \Big)
    \right|_{\chi = \chi_{\rm max}}
    \,.
\la {eq:Qmax}
\end {equation}
In more explicit form, the two terms in $Q$ are
\begin {eqnarray}
   \Big( \chi_\ij, S_\ij \Big)
    &=&
    -\beta^2 \sum^\ffhc_a
    \int_\p
	f_0(\p) [1\pm f_0(\p)] \>
	q^a \chi^a(|\p|) \,,
\label {eq:Sij}
\\
\noalign {\hbox {and}}
    \Big( \chi_\ij, {\cal C} \chi_\ij \Big)
    &=&
    \frac {\beta^3}8
    \sum^\ffhc_{abcd}
    \int_{\p,\k,\p',\k'}
	\left| M^{ab}_{cd}(p,k,p',k') \right|^2 \>
	(2\pi)^4 \, \delta^4 (p+k-p'-k')
\nonumber\\ && \kern 0.6in {} \times
	f^a_0(\p) \, f^b_0(\k) \, [1{\pm}f^c_0(\p')] \, [1{\pm}f^d_0(\k')]
\nonumber\\ && \kern 0.6in {} \times
	\Bigl[
	    \chi^a_\ij(\p) + \chi^b_\ij(\k) -
	    \chi^c_\ij(\p') - \chi^d_\ij(\k')
	\Bigr]^2 \,.
\la {eq:Q2}
\end {eqnarray}
The sum is over all scattering processes in the plasma
taking species $a$ and $b$ into species $c$ and $d$.
We have used crossing symmetry of the scattering amplitudes
to write the above expression for matrix elements of $\cal C$
in a form which makes it apparent that ${\cal C}$ is a
positive semi-definite operator.
The overall factor of 1/8 compensates for the eight times
a given process appears in the multiple sum over species
when all the particles are distinct (due to relabeling
$a\leftrightarrow b$, $c\leftrightarrow d$, and/or $ab\leftrightarrow cd$),
and supplies the appropriate symmetry factor in cases
where some or all of the particles are identical.
For our later discussion it will be important to note that
${\cal C}$ is non-diagonal in the basis of species when there
are $2\leftrightarrow 2$ processes involving more than one species type.

After solving the linearized Boltzmann equation
in the particular channel of interest,
by maximizing $Q[\chi]$,
the associated transport coefficient may be determined by inserting
the resulting distribution function
[given by Eqs.~(\ref {eq:f1}) and (\ref {eq:fa1})]
into the stress-energy tensor (\ref {eq:kineticT}) or
appropriate conserved current (\ref {eq:kineticJ}), and
comparing with the corresponding constitutive relation
[Eq.~(\ref {eq:Tij}), (\ref {eq:jiEM}), or (\ref {eq:jia})].
In each case,
the integral over the distribution function which defines the flux
({\em i.e.}, the stress tensor $T_{ij}$ or a spatial current $j^\alpha_i$)
reduces to the inner product $(\chi_\ij, S_\ij)$.
Consequently,
the actual value of each transport coefficient turns out to be
trivially related to the maximum value (\ref {eq:Qmax})
of the functional $Q$ in the corresponding channel.
Explicitly,
\begin {eqnarray}
    \sigma
    &=&
    {\textstyle {2\over 3}} \,
    Q_{\rm max} \Bigr|_{\ell = 1, \; q = q_{\rm EM}} \,,
\la {eq:sigmaQ}
\\[5pt]
    D_\alpha
    &=&
    {\textstyle {2\over 3}} \,
    Q_{\rm max} \Bigr|_{\ell = 1, \; q = q_\alpha}
    \left( {\partial n_\alpha \over \partial\mu_\alpha} \right)^{-1} ,
\la {eq:DQ}
\\[5pt]
    \eta
    &=&
    {\textstyle {2\over 15}} \,
    Q_{\rm max} \Bigr|_{\ell = 2, \; q = |\p|} \,.
\la {eq:etaQ}
\end {eqnarray}
The thermodynamic derivative appearing in (\ref {eq:DQ}) is
the charge susceptibility,
\begin {eqnarray}
    \suscept_\alpha
    \equiv
    {\partial n_\alpha \over \partial \mu_\alpha}
    &=&
    \sum^\ffhc_a \> (q_\alpha^a)^2 \int_\p \beta \, 
	f^a_0(\p) [1 \pm f^a_0(\p) ]
\nonumber\\
    &=&
    {\textstyle {1 \over 12}} \, T^2 \>
    \sum^\ffhc_a \, \lambda_a \, (q_\alpha^a)^2 \,,
\la {eq:dndmu}
\end {eqnarray}
where, once again,
$\lambda_a$ is 1 for fermions and 2 for bosons, and we have specialized to 
the ultra-relativistic limit.

\section {Collision integrals}
\la {sec:leading-log}

\begin{figure}[t]
\centerline{\epsfxsize=5in\epsfbox{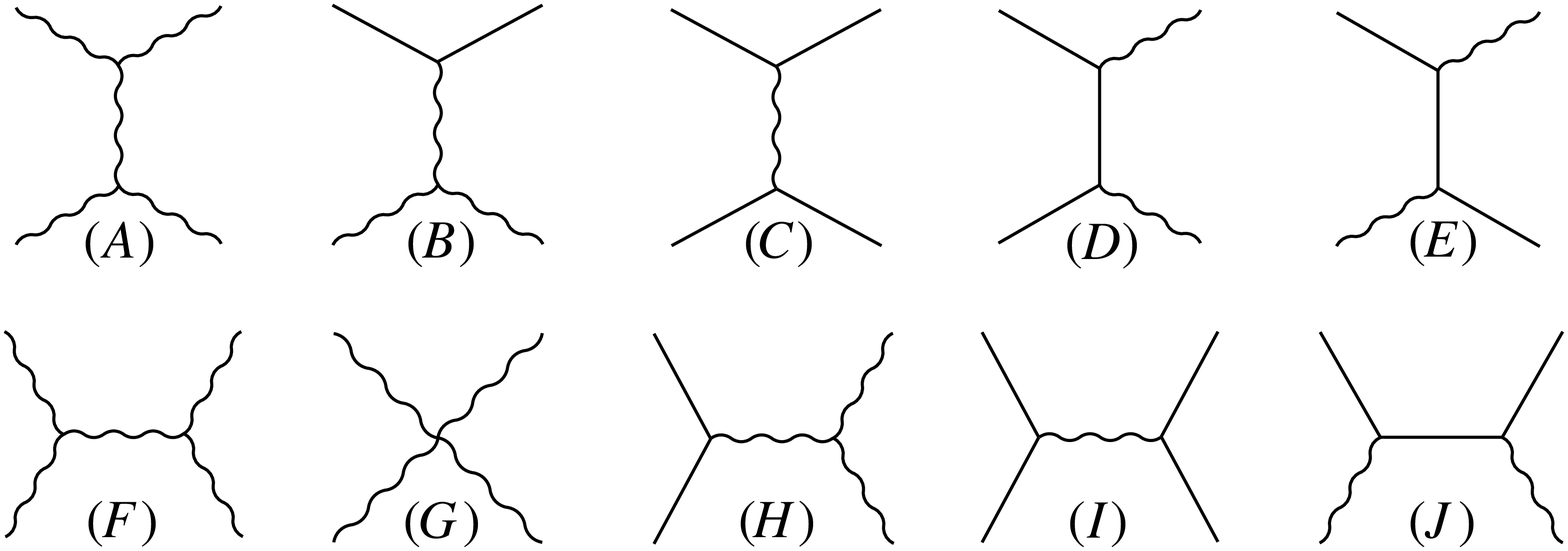}}
\vspace{0.2in}
\caption{%
    \la{fig:diagrams}
    Leading-order diagrams for all $2\leftrightarrow 2$ particle
    scattering processes in a gauge theory with fermions.
    Solid lines denote fermions and wiggly lines are gauge bosons.
    Time may be regarded as running horizontally, either way, and
    so a diagram such as $(D)$ represents both
    $f \bar f \to gg$ and $gg \to f \bar f$.
    The diagrams of the first row [$(A)$--$(E)$]
    contribute to the leading log transport coefficients,
    while the diagrams of the second row [$(F)$--$(J)$],
    and all interference terms, do not.
}%
\end{figure}

The full set of scattering processes which contribute at leading order
in a theory with gauge and fermionic degrees of freedom are shown in
Fig.~\ref{fig:diagrams}.
Some of these processes yield matrix elements which become singular as
the momentum transfer ({\em i.e.}, Mandelstam $t$ or $u$) goes to zero.  
For instance, in a vector-like theory, the matrix element for
gauge boson exchange between fermions
[diagram $(C)$] behaves, for non-identical fermions, as
\begin{equation}
{\cal M}^2_{{\rm diagram \;} C} \propto \frac{s^2 + u^2}{t^2}
	\mathop{\longrightarrow}\limits_{\p \to \p'}
	\; 2 \, \frac{[(p+k)_\mu (p+k)^\mu]^2}
	{[(p-p')_\nu (p-p')^\nu]^2} \, ,
\end{equation}
which diverges as the inverse fourth power of the momentum transfer
as $\p$ approaches $\p'$.
Phase space only partially compensates, resulting in a
cross section which is quadratically divergent at small $\p-\p'$.
Similarly, the annihilation diagram $(D)$
has ${\cal M}^2 \propto (u/t) + (t/u)$,
which leads to a logarithmically IR divergent scattering cross section. 
However,
we are not directly interested in the total scattering cross section;
we need to know the size of the contribution to
$\big( \chi_\ij, {\cal C} \chi_\ij \big)$
in the channels relevant to transport coefficients.
As we shall review, these transport collision integrals
can be less singular than the total scattering rate.

\subsection{Kinematics}

It is
convenient to arrange the phase space integrations so that
the transfer momentum is explicitly an integration variable.
This will make it easy to isolate the contribution from the
potentially dangerous small momentum exchange region.
We choose to label the outgoing particles so that any
infrared singularity in (a given term of) the square of
the amplitude
$|{\cal M}|^2$ occurs only when $(p'{-}p)^2 \rightarrow 0$.%
\footnote
    {%
    There is one case where this is impossible,
    namely, scattering between identical fermions,
    where the interference term between outgoing leg
    assignments in diagram $(C)$ makes a contribution to the
    matrix element ${\cal M}^2 \propto s^2/ut$, which is divergent
    for both $t\to0$ and $u\to0$.
    As will be discussed shortly, this interference does not contribute at
    leading-log level.  Regardless, one could also put this case in
    the desired form by using
    $s=-u-t$ and rewriting the matrix element (squared)
    as $(s/t) + (s/u)$, so that each piece is now singular in only
    one momentum region.  Diagram $(A)$ apparently has the same problem; 
    but when one sums all $gg \rightarrow gg$ processes (only the sum 
    is gauge invariant) one finds 
    ${\cal M}^2 \propto ( 3 - su/t^2 - st/u^2 - tu/s^2 )$, so there is
    no problem.
    }  %
In the collision integral (\ref {eq:Q2})
it is convenient to use the spatial $\delta$ function to perform the
$\k'$ integration, and to shift the $\p'$ integration into an
integration over $\p'{-}\p \equiv \q$.
We may write the angular integrals
in spherical coordinates with $\q$ as the $z$ axis and choose the $x$
axis so $\p$ lies in the $x$-$z$ plane.
This yields
\begin{eqnarray}
    \Big( \chi_\ij, {\cal C} \chi_\ij \Big)
    &=&
    \frac{\beta^3}{(4\pi)^6}
    \sum^\ffhc_{abcd}
    \int_{0}^{\infty} q^2 dq \> p^2 dp \> k^2 dk 
	\int_{-1}^{1} d \cos\theta_{pq} \> 
	d\cos\theta_{kq}
	\int_0^{2\pi} d\phi
\nonumber \\ && \hspace {0.9cm} {}\times
	\frac{1}{p\,k\,p'\,k'} \> \left|{\cal M}^{ab}_{cd}\right|^2 \>
	\delta(p{+}k{-}p' {-} k') \>
	f^a_0(p) \, f^b_0(k) \, [1 \pm f^c_0(p')] \, [1 \pm f^d_0(k')]
\nonumber \\ && \hspace {0.9cm} {}\times
	\left[
	    \chi^a_\ij(\p) + \chi^b_\ij(\k) - \chi^c_\ij(\p') - \chi^d_\ij(\k')
	\right]^2
	 \, ,
\end{eqnarray}
where here and henceforth, $p$, $k$, and $q$ denote the magnitudes of the
corresponding three-momenta (not the associated 4-momenta),
$p' \equiv |\q+\p|$ and $k' \equiv |\k-\q|$ are the magnitudes of the 
outgoing momenta,
$\phi$ is the azimuthal angle of $\k$ (and $\k'$)
[{\em i.e.}, the angle between the $\p$-$\q$ plane and the $\k$-$\q$ plane],
and $\theta_{pq}$ is the plasma frame angle between $\p$ and $\q$,
$\cos\theta_{pq} \equiv \hat\p \cdot \hat\q$, {\em etc}.

Following Baym {\it et~al.}~\cite{BMPRa}, it is convenient to
introduce a dummy
integration variable $\omega$, constrained by a $\delta$ function to equal
the energy transfer $p' - p$,
so that
\begin{equation}
    \delta(p+k-p' - k')
    =
    \int_{-\infty}^\infty d\omega \>
    \delta(\omega + p - p') \, \delta(\omega-k+k') \, .
\end{equation}
Evaluating $p'=|\p+\q|$
in terms of $p$, $q$, and $\cos \theta_{pq}$, and 
defining $t = \omega^2 - q^2$ (which is the usual Mandelstam variable),
one finds
\begin{eqnarray}
    \delta(\omega+p-p')
    &=&
    \frac{p'}{pq} \>
    \delta\biggl( \cos \theta_{pq} - \frac{\omega}{q} 
	- \frac{t}{2pq} \biggr) \,
    \Theta(\omega+p) \,,
\\
    \delta(\omega-k+k')
    &=&
    \frac{k'}{kq} \>
    \delta\biggl( \cos \theta_{kq} - \frac{\omega}{q} 
	+ \frac{t}{2kq} \biggr) \,
    \Theta(k-\omega) \,,
\end{eqnarray}
where $\Theta(z)$ is the unit step function.
The $\cos\theta$ integrals may now be trivially performed and yield one
provided $p > \half(q-\omega)$, $k>\half(q+\omega)$, and
$|\omega|<q$; otherwise the argument of a $\delta$ function has no
zero for any $|\cos \theta| \leq 1$.
The remaining integrals are
\begin{eqnarray}
    \Big( \chi_\ij, {\cal C} \chi_\ij \Big)
    &=&
    \frac{\beta^3}{(4\pi)^6}
    \sum^\ffhc_{abcd}
    \int_0^{\infty} dq
    \int_{-q}^q d\omega 
    \int_{\frac{q-\omega}{2}}^\infty dp 
    \int_{\frac{q+\omega}{2}}^\infty dk
    \int_0^{2\pi} d\phi
\nonumber \\ && \hspace {1cm} {}\times
	\left|{\cal M}^{ab}_{cd}\right|^2 \>
	f^a_0(p) \, f^b_0(k) \, [1 \pm f^c_0(p')] \, [1 \pm f^d_0(k')]
\nonumber \\ && \hspace {1cm} {}\times
	\left[
	    \chi^a_\ij(\p) + \chi^b_\ij(\k) - \chi^c_\ij(\p') - \chi^d_\ij(\k')
	\right]^2
	 \, ,
\la {eq:Q3}
\end{eqnarray}
with $p'=p+\omega$, and $k'=k-\omega$.
For evaluating the final factor of (\ref {eq:Q3}),
note that
\begin{equation}
    I_\ij(\hat\p) \, I_\ij(\hat\k)
    =
	P_\ell(\cos\theta_{pk}) \,,
\la{eq:contract_I}
\end{equation}
where $P_\ell$ is the $\ell$'th Legendre polynomial.
We will therefore need expressions for the angles between the momenta of
all species,
as well as the remaining Mandelstam variables $s$ and $u$, which 
appear in ${\cal M}^2$.
They are
\begin{eqnarray}
\cos \theta_{pq} & = & \frac{\omega}{q} + \frac{t}{2pq} 
	\, , \hspace{1.47in} \;
\cos \theta_{p'q} = \frac{\omega}{q} - \frac{t}{2p'q} \, , \\
\cos \theta_{kq} & = & \frac{\omega}{q} - \frac{t}{2kq} 
	\, , \hspace{1.5in}
\cos \theta_{k'q} = \frac{\omega}{q} + \frac{t}{2k'q} \, , \\
\cos \theta_{pp'} & = & 1 + \frac{t}{2pp'} \, ,\hspace{1.52in}
\cos \theta_{kk'} = 1 + \frac{t}{2kk'} \, , \\
\cos \theta_{pk}\: \: & = &  \cos \theta_{pq} \: \cos \theta_{kq} \: 
	 + \sin \theta_{pq} \: \sin \theta_{kq} \: \cos \phi \, , 
\la{eq:mixed_angle} \\
\cos \theta_{pk'} \: & = & \cos \theta_{pq} \: \cos \theta_{k'q} 
	 + \sin \theta_{pq} \: \sin \theta_{k'q} \cos \phi \, , \\
\cos \theta_{p'k} \: & = & \cos \theta_{p'q} \cos \theta_{kq} \:
	 + \sin \theta_{p'q} \sin \theta_{kq} \: \cos \phi \, , \\
\cos \theta_{p'k'} & = & \cos \theta_{p'q} \cos \theta_{k'q} 
	 + \sin \theta_{p'q} \sin \theta_{k'q} \cos \phi \, , 
\end {eqnarray}
and
\begin {eqnarray}
\la{eq:s_equals}
    s &=&
      2 p k \, (1 - \cos\theta_{pk})
=
        \frac{-t}{2q^2}
	\left\{
	(p{+}p')(k{+}k')+q^2 
	- \cos\phi \,
	    \sqrt{\textstyle
	    \left(4pp'+t\right)\left(4kk'+t\right) } \,
	\right\} ,
\\
u & = & -t -s \, .
\end{eqnarray}

\subsection{Leading-log simplifications}

To compute leading-log transport coefficients, it will be sufficient to
extract the small $q$ contribution to the collision integral
(\ref {eq:Q3}).
For small $q$ but generic $p$ and $k$
(so that $q\ll T$, $q\ll p$, and $q\ll k$), one has
\begin{equation}
    \frac{-s}{t} \simeq \frac{u}{t} \simeq \frac{2 p k }{q^2}  \, (1-\cos \phi)
	\,.  
\end{equation}
The $\omega$ integral is restricted to the range $-q \le \omega \le q$
and will be dominated by $\omega/q$ of order one.
Hence, the small $q$ integration region in 
a diagram where ${\cal M}^2 \propto (s^2 \mbox{ or } u^2)/t^2$ naively
behaves like $\int dq/q^3$, a quadratic infrared divergence; 
while for a diagram with 
${\cal M}^2 \propto st/t^2$ (or $ut/t^2$ or $s^2/tu$) the small $q$
integration region naively behaves like $\int dq/q$, a logarithmic divergence.
These estimates are inadequate, however, if species $a$ and $c$
are identical, and species $b$ and $d$ are also identical ---
that is, when both incident particles undergo
small angle scattering without changing their species types.
In this case, since $\p'-\p=\q$ is small, one has 
\begin{equation}
    \chi^a_\ij(\p) - \chi^a_\ij(\p')
    =
    -\q \cdot \naBla \chi^a_\ij( \p ) + O(q^2)\,,
\la{eq:cancel}
\end{equation}
and similarly for
$ \chi^b_\ij(\k)-\chi^b_\ij(\k')\, $.
Therefore,
the $[\chi^a {+} \chi^b {-} \chi^c {-} \chi^d]^2$ factor in the collision
integral (\ref {eq:Q3}) will contribute
a factor of $q^2$ to the integrand,
softening the small $q$ divergence.

This $q\to0$ cancellation is operative for diagrams $(A)$, $(B)$, and $(C)$,
and converts the naive estimate of a quadratic infrared divergence
into a merely logarithmic divergence.
It also converts interference terms involving these diagrams,
which were naively log divergent, into finite contributions.
The cancellation does
{\em not} occur for diagrams $(D)$ and $(E)$ which involve 
a change of species.
For these diagrams, the naive estimate of an infrared log divergent
result is correct.
Interference terms involving these diagrams, and the remaining
diagrams $(F)$--$(J)$, are infrared finite from the outset.
Hence, (the squares of) all diagrams in the first row in
Fig.~\ref{fig:diagrams} lead to logarithmic IR divergences
in Eq.~(\ref{eq:Q3}), 
while diagrams in the second row, and all interference terms, do not.

All of these logarithmic divergences become convergent when one includes 
the self-energies appearing on exchange lines in the diagrams.  The
self-energies are all of order $g^2 T^2$.
For small $q \ll T$, the $g^2 T^2$ part of the self-energies
are known as the hard thermal loop (HTL) self-energies \cite{HTL1,HTL2}.
An $O(g^2 T^2)$ self-energy correction on a 
propagator is important when the exchange momentum squared becomes of 
order $g^2 T^2$, which means when $q \sim gT$.  In every case, the
inclusion of the self-energy reduces the size of the 
matrix element and serves to
cut off the log divergence in the infrared.  For the case of the gauge
boson propagator, relevant in diagrams $(A)$--$(C)$, this is discussed
in \cite{BMPRa}.  The demonstration that the log is cut off 
for diagrams $(D)$ and $(E)$ has apparently 
not appeared in the previous literature, though the self-energy for the
fermion line is well known \cite{Weldon2}.  Our analysis shows that the
self-energy on the fermion line is sufficient to cut off this IR
divergence as well \cite{all-log}.  In the current paper we will not
discuss this issue in detail; nor will we treat carefully the momentum
region $q \gsim T$, where the small $q$ approximations which led to
the conclusion that there is a log divergence break down,
cutting off the log from above.
Instead we will make a {\em leading-log} treatment, which means
that we will extract the {\em coefficient} of the logarithmic
divergence.  This permits us to simultaneously take $q$ to be small, $q
\ll T$, $q \ll p$, $q \ll k$, allowing certain kinematic
simplifications, while simultaneously neglecting self-energy corrections in
determining the matrix elements.  This approximation has been customary
in almost all work in the field; in fact we are not aware of any paper
which correctly goes beyond this approximation when computing a
transport coefficient in a relativistic gauge theory.  In a companion paper,
we will treat the problem to full leading order in $g$ \cite{all-log}.

\subsection{Diagrams $(A)$, $(B)$, and $(C)$}

\begin{figure}[t]
\centerline{\epsfxsize=1.5in\epsfbox{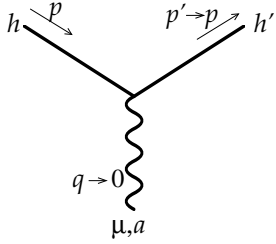}}
\vspace{0.2in}
\caption{%
    \la{fig:vertex}
    A generic vertex from diagrams ($A$)--($C$) of Fig.\ \ref{fig:diagrams},
    to be analyzed in the soft exchange limit.  In this figure, the
    solid line denotes
    any sort of particle (e.g., a gauge boson or fermion) that is being
    scattered, and the wavy line represents the exchanged gauge boson.
}%
\end{figure}

Consider the gauge boson exchange diagrams $(A)$, $(B)$, and $(C)$.
These represent $ab \leftrightarrow ab$ processes where,
in the relevant small $q$ regime,
the incoming and outgoing lines with nearly the same momenta
are the same species type.
The near cancellation (\ref{eq:cancel}) thus implies that
the $\chi$ factors will contribute an explicit $q^2$
for small $q$.
We may therefore use leading small $q$ approximations everywhere else,
together with the first {\em nontrivial} small $q$ approximation for
the $\chi_\ij$ factors.
As a result, the various forms which appear in the square of the
matrix elements for the different diagrams $(A)$--$(C)$
all become the same to leading order.
In fact, in the soft exchange ($q\to0$) limit, the vertices in diagrams
($A$)-($C$) take on a universal form
that depends only on the color charge of the particle that is scattering,
be it a gluon or quark (or scalar).
Such a vertex is depicted generically in Fig.\ \ref{fig:vertex}, and the
associated vertex factor in the $q\to0$ limit is%
\footnote{
   One may alternatively work directly with the full matrix elements and
   observe that their structure becomes identical in the $q\to0$ limit because
   $
    (s^2 {+} u^2)/(2 t^2) \simeq {-su}/{t^2} \simeq {s^2}/{t^2} 
	\simeq (4 p^2 k^2/q^4) \left( 1 - \cos \phi \right)^2
   $.
}
\begin {equation}
   2 p^\mu g \, t^a \, \delta_{hh'} ,
\end {equation}
where $t^a$ is the color generator in the representation of the scattering
particle, and $h$ and $h'$ are the ingoing and
outgoing helicities of that particle.
(The $2 p$ in the factor $2p^\mu = 2 p v^\mu$ 
is a consequence of using relativistic normalization
for matrix elements.)  With this $q\to0$ Feynman rule, it is trivial
to evaluate the matrix elements for
the $ab\leftrightarrow ab$ scattering processes of
Figs. ($A$)--($C$):
\begin {equation}
    \left| {\cal M}^{ab}_{ab} \right|^2_{\rm leading-log}
    =
    A_{ab} \, {s^2 \over t^2} \,,
\end {equation}
where the coefficient $A_{ab}$ depends on the gauge coupling
and representations of the two species,%
\footnote
    {%
    If $a{=}b$, then the matrix element has a second $s^2/u^2$ term
    arising from the interchange of the outgoing lines.
    Swapping the $\p'$ and $\k'$ labels effectively makes the
    resulting matrix element twice as large.%
    \label{foot:Melement}
    }
\begin {equation}
    A_{ab} \equiv 4 \, \da \, T_{{\rm R}_a} \, T_{{\rm R}_b} \> g^4 \,.
\end {equation}
Strictly speaking, this gives the square of the amplitude summed over
incoming and outgoing gauge group indices and outgoing spins or
helicities, but not summed over incoming spins or helicities,
which are considered part of the species label.
The corresponding contribution to the collision integral becomes
\begin {eqnarray}
    \Big( \chi_\ij, {\cal C} \chi_\ij \Big)^{{\rm diagrams}\;(A{-}C)}
    _{\rm leading-log}
    &=&
    \sum^\ffh_{ab}
    \frac {2A_{ab} \,\beta^3}{(4 \pi)^6}
    \int_0^\infty dq
    \int_{-q}^q d\omega
    \int_0^\infty dk
    \int_0^\infty dp
    \int_0^{2\pi} d\phi \;
	(1{-}\cos \phi)^2
\nonumber\\ && \qquad {} \times
	\frac{4 p^2k^2}{q^4} \,
	f_0^a(p) \, [1\pm f_0^a(p)] \, f_0^b(k) \, [1\pm f_0^b(k)]
\nonumber\\ && \qquad {} \times
	\left[ \Big( \chi^a_\ij(\p) - \chi^a_\ij(\p') \Big)
	    +  \Big( \chi^b_\ij(\k) - \chi^b_\ij(\k') \Big) \right]^2 .
\label {eq:ABC1}
\end{eqnarray}
As in Eq.~(\ref {eq:approx_D}),
the species sums run over all helicities and types of excitations,
counting anti-particles separately from particles.
The 2 next to $A_{ab}$ arises because the sum in Eq.~(\ref{eq:Q2})
separately counts both $ab\rightarrow ab$ and $ab \rightarrow ba$.%
\footnote
    {%
    Unless $a{=}b$, in which case the extra factor of
    two comes from the matrix element, as noted in
    the previous
    footnote. 
    }%

The leading small $q$ approximation (\ref {eq:cancel})
to the difference of $\chi$ factors,
for either $\ell = 1$ or $\ell = 2$,
has the explicit form
\begin {mathletters}
\begin {eqnarray}
    \chi^a_{i}(\p) - \chi^a_{i}(\p')
    &=&
    - \omega \, I_{i}(\hat \p) \, \chi^a(p)'
    + \left( \omega \, \hat p_i - q_i \right) { \chi^a(p) \over p}
    + O(q^2) \,,
\\
    \chi^a_{ij}(\p) - \chi^a_{ij}(\p')
    &=&
    - \omega \, I_{ij}(\hat \p) \, \chi^a(p)'
    + {\textstyle \sqrt{3\over 2}}
	\left(
	    2\, \omega\, \hat p_i \hat p_j\, 
	    - q_i \, \hat p_j
	    - q_j \, \hat p_i
	\right) {\chi^a(p) \over p}
    + O(q^2) \,,
\end {eqnarray}%
\label {eq:chidiff}%
\end {mathletters}%
where $\chi^a(p)'$ means $d\chi^a(p)/dp$.
Expressions for $\chi^b_\ij(\k) - \chi^b_\ij(\k')$
are the same except for replacing $\p$ by $\k$,
and changing the overall sign.
For either case (and in fact, for any $\ell$), 
\begin{equation}
    \Bigl[ \chi^a_\ij(\p) - \chi^a_\ij(\p') \Bigr]^2
    =
    \omega^2 \, \left[ \chi^a(p)'\right]^2
    + \half \, {\ell(\ell{+}1)} \, \frac{q^2 - \omega^2}{p^2} \,
    \left[ \chi^a(p) \right]^2
    + O(q^3) \,.
\label {eq:chidiffsq}
\end{equation}

When expanding the last factor of Eq.~(\ref {eq:ABC1}),
there are two types of contributions to examine: those
involving two $\chi^a$ or two $\chi^b$ factors,
for which one may use the expression (\ref {eq:chidiffsq}),
and the cross-contributions with one $\chi^a$ and one $\chi^b$.
We will consider the cross-contributions first.
As noted above,
the explicit $q$ or $\omega$ appearing in the difference (\ref {eq:chidiff})
softens the small $q$ behavior to a logarithmic divergence,
so in all other factors one may work to leading order in $\q$.
In particular one may approximate 
\begin{eqnarray}
\cos \theta_{pq} & \simeq & \cos \theta_{kq} \simeq
	\cos \theta_{p'q} \simeq \cos \theta_{k'q} 
	\simeq {\omega \over q} \, , 
\la{eq:ll_angle1} \\
\cos \theta_{pk} & \simeq & \cos \theta_{p'k} \simeq 
	\cos \theta_{pk'} \simeq \cos \theta_{p'k'}
	\simeq \frac{\omega^2}{q^2} + \frac{q^2 - \omega^2}{q^2}
	\cos \phi \, .
\la{eq:ll_angle2}
\end{eqnarray} 
Explicitly carrying out the
$d\omega$ and $d\phi$ integrations, 
with the $(1-\cos \phi)^2$ factor from the matrix element included,
one finds that all $\chi^a \chi^b$ cross terms vanish in case
of $\ell = 2$ (or higher), but not for $\ell = 1$.
In the $\ell=1$ channels, however,
we will only be interested in the diffusion of charge conjugation
(C) odd quantum numbers (electric charge or various fermionic numbers like
baryon number).
In our high temperature regime,
the equilibrium state may be regarded as C (or CP) invariant.
Consequently, C (or CP) symmetry ensures that 
particles and anti-particles will have opposite
departures from equilibrium,
$\chi^{\overline{a}} = - \chi^a$
(where $\overline a$ denotes the anti-particle of species $a$).
Hence, the sign of the cross term will be different for scatterings
from fermions versus anti-fermions, so that the two contributions
will cancel in the sum over species.
When the scattering involves a gauge boson on one or both lines,
then C symmetry ensures that the departure from equilibrium for
the gauge boson is zero, so again there is no cross term.
For the same reason, gauge-boson---gauge-boson scattering [diagram $(A)$]
plays no role for conductivity or diffusion.

In either case, what remains are only the terms with two $\chi^a$
or two $\chi^b$ factors.
After using (\ref {eq:chidiffsq})
(or the same relation with
$a\leftrightarrow b$ and $p \leftrightarrow k$),
the $\omega$ and $\phi$ integrals are simple,
and the $k$ integral can also be performed using
\begin{equation}
    \int_0^\infty dk \> k^2 f_0^b(k) \, [1\pm f_0^b(k)]
    =
    \lambda_b \, T^3 \, \frac{\pi^2}{6} \, ,
\label {eq:b(1+b)}
\end{equation}
where $\lambda_b = 2$ if species $b$ is bosonic, and 1 if it is fermionic.
The result is
\begin{eqnarray}
    &&
    \Big( \chi_\ij, {\cal C} \chi_\ij \Big)^{{\rm diagrams}\;(A{-}C)}
    _{\rm leading-log} \qquad
\nonumber \\ && \qquad {} =
    \sum^\ffh_{ab}
    \frac{A_{ab}}{2^{9} \, 3 \pi^3}\int_{gT}^{T} {dq \over q}
    \!\int_0^\infty \!\! dp \,
    \biggl[
	\lambda_b \, f_0^a(p) \,[1 \pm f_0^a(p)] \!
	\left(
	    p^2 \Big[\chi^a(p)'\Big]^2 + \ell(\ell{+}1) \,\chi^a(p)^2
	\right)
    + (a \leftrightarrow b) \biggr]
\nonumber \\ && \qquad {} =
    \sum^\ffh_{ab}
    \frac{A_{ab}}{2^{8} \, 3 \pi^3}\int_{gT}^{T} {dq \over q}
    \int_0^\infty \! dp \>
	\lambda_b \, f_0^a(p) \,[1 \pm f_0^a(p)]
	\left(
	    p^2 \Big[\chi^a(p)'\Big]^2 + \ell(\ell{+}1) \,\chi^a(p)^2
	\right) .
\la{eq:B_is}
\end{eqnarray}

In the $q$ integration, the upper cutoff is $q \sim T$, where
the small $q$ treatment breaks down.  The lower cutoff occurs because we 
have not included the gauge boson's hard thermal loop self-energy in
computing the matrix element.  Inclusion of the self-energy makes the
matrix element smaller than the vacuum amplitude and cuts off the $q$
integration in the infrared.%
\footnote
    {%
    This is true for both longitudinal and transverse parts
    of the exchanged gauge boson propagator
    \cite{BMPRa}.
    }
Hence, in a leading log treatment one may simply replace the entire
$q$ integral by $\log g^{-1}$, and thereby reduce these
collision integral contributions to a single one-dimensional integral.

\subsection{Diagrams $(D)$ and $(E)$}

We begin with the annihilation diagram $(D)$.
The matrix element squared for a fermion of species $f$
to annihilate with an anti-fermion of opposite helicity
and produce two gauge bosons, $f \f \to gg$,
summed over initial and final gauge group indices and
gauge boson spins, is
\begin {equation}
    \left| {\cal M}^{f\f}_{gg} \right|^2_{\rm leading-log}
    =
    A_{f} \left( {u \over t} + {t \over u} \right) \,,
\label {eq:Mffgg}
\end {equation}
with
\begin {equation}
    A_{f} \equiv 4 \, \da \, T_{{\rm R}_f} \, C_{{\rm R}_f} \> g^4 \,.
\end {equation}
Interchanging labels on the outgoing legs turns $(t/u)$ into $(u/t)$, 
and so one may keep just the $u/t$ part of the matrix element
and multiply the result by two [which effectively cancels part of the
overall 1/8 symmetry factor which appears in Eq.~(\ref {eq:Q2})].

Since the degree of divergence is at most logarithmic, we may
immediately make all available small $q$ approximations.  Namely, we may 
take the limits of the $p$ and $k$ integrations to be zero,
take $f_0(p{+}\omega) \simeq f_0(p)$
and similarly $f_0(k{-}\omega)\simeq f_0(k)$,
approximate $\cos \theta_{pp'} \simeq \cos \theta_{kk'} \simeq 1$,
and use Eqs.~(\ref{eq:ll_angle1}) and (\ref{eq:ll_angle2}) for the
various angles.
The matrix element, at leading order in small $q$, is just
\begin{equation}
    \frac{u}{t} \simeq \frac{2 k p }{q^2} \, (1 - \cos \phi) \,.
\end{equation}
Making these approximations gives the following contribution
to the collision integral (\ref{eq:Q2}),
\begin{eqnarray}
    \Big( \chi_\ij, {\cal C} \chi_\ij \Big)^{{\rm diagram}\;(D)}
    _{\rm leading-log}
    &=&
    \sum^\fh_{f}
    \frac{8 \, A_{f} \,\beta^3}{(4\pi)^6}
    \int_0^\infty dq
    \int_{-q}^q d\omega
    \int_0^\infty dk
    \int_0^\infty dp
    \int_0^{2\pi} d\phi \; (1 - \cos \phi)
\nonumber\\ && \qquad {} \times
	\frac{2 pk}{q^2} \,
	f_0^f(p) \, f_0^\f(k) \, [1+f_0^g(p)] \, [1+f_0^g(k)]
\nonumber\\ && \qquad {} \times
	\Big\{
	    \Bigl[\chi^f(p)-\chi^g(p) \Bigr]^2 +
	    \Bigl[ \chi^{\f}(k) -\chi^g(k) \Bigr]^2
\nonumber \\ && \qquad \qquad {}
	    + 2 P_\ell(\cos \theta_{pk}) \,
	    \Bigl[ \chi^f(p) -\chi^g(p) \Bigr]
	    \Bigl[\chi^{\f}(k) - \chi^g(k) \Bigr]
	\Big\} \, .
\end{eqnarray}
Here $\chi^f$ is the departure from equilibrium for the fermion,
$\chi^{\f}$ is for its anti-particle, and $\chi^g$
is for the gauge boson.
The distribution function
$f_0^f$ is the equilibrium Fermi distribution
while $f_0^g$ is the equilibrium Bose distribution.
The sum runs over all fermion species and helicities, but not over
anti-particles.
The factor of $8$ next to $A_{f}$ is the $2$ from the two pieces of the
matrix element, times the
4 ways Eq.~(\ref{eq:Q2}) counts this diagram
($f\f \rightarrow gg,$ $\f f \rightarrow gg$, $gg \rightarrow f\f$, 
$gg \rightarrow \f f$).

Using Eq.~(\ref{eq:ll_angle2}) for $\theta_{pk}$,
one may easily check that
\begin{equation}
\int_{-q}^q d\omega \int_0^{2\pi} d\phi \;
	(1 - \cos \phi) \, P_\ell(\cos \theta_{pk}) = 0
\la{eq:approx_pk}
\end{equation}
for $\ell=1$ and $\ell=2$ (and in fact, all $\ell>0$).
Therefore the cross term involving both $\chi^f$ and $\chi^{\f}$
vanishes.  For the remaining terms, the $\omega$ and $\phi$
integrals are trivial.
For the term involving $[\chi^{\f}(k) {-} \chi^g(k)]^2$,
one may use the symmetry of the integration region to exchange $p$ and $k$.
After performing the $k$ integration with
\begin{equation}
    \int_0^\infty dk \> k \, f_0^f(k) \, [1 + f_0^g(k)]
    = \frac{\pi^2 \, T^2}{8} \, ,
\end{equation}
one finds
\begin{eqnarray}
    \Big( \chi_\ij, {\cal C} \chi_\ij \Big)^{{\rm diagram}\;(D)}
    _{\rm leading-log} 
    &=& \sum^\fh_{f}
    \frac{A_{f} \, \beta}{2^{9} \, \pi^3}
    \int_{gT}^{T}
    \frac{dq}{q}
    \int_0^\infty dp \> p \, f_0^f(p) \, [1+f_0^g(p)]
\nonumber \\ && \hspace{1in}  {} \times
	\left\{
	    \Big[ \chi^f(p) - \chi^g(p) \Big]^2 +
	    \Big[ \chi^{\f}(p) - \chi^g(p) \Big]^2
	\right\} \, .
\la{eq:D_is}
\end{eqnarray}

Once again,
the limits on the $q$ integration show
where the approximations we have used break down.  The small $q$
approximation breaks down for $q \gsim T$.
Approximating
the matrix element by the vacuum amplitude, without a thermal self-energy
insertion on the internal fermion propagator, 
is invalid for $q \lsim gT$.
Below this scale the matrix element is smaller --- parametrically smaller
for $q \ll gT$.
Our approximations are an over-estimate for $q \lsim gT$, and the
integral is cut off.  For a leading log treatment, one may simply replace
the $q$ integration with $\log(T/gT)=\log g^{-1}$, which reduces
the collision integral contribution (\ref {eq:D_is})
to a single integral over a quadratic form in the departures from equilibrium.

Finally, the Compton scattering diagram $(E)$
differs only slightly from the annihilation diagram $(D)$.
The matrix element for diagram $(E)$ is $-(s/t)$ rather than $(u/t)$,
but at leading order in small $q$ these are equivalent.
The sign of each $\chi$ associated with excitations with momentum $\k$
is opposite,
and the fermion departures from equilibrium are now either both
$\chi^f$ or both $\chi^{\f}$,
rather than one of each.
Summing both particle and anti-particle cases,
and exploiting the fact that the $p$--$k$ cross-term cancels,
one finds that the leading log contribution of diagram $(E)$ is
identical to that of diagram $(D)$,
\begin{equation}
    \Big( \chi_\ij, {\cal C} \chi_\ij \Big)^{{\rm diagram}\;(E)}
    _{\rm leading-log}
    =
    \Big( \chi_\ij, {\cal C} \chi_\ij \Big)^{{\rm diagram}\;(D)}
    _{\rm leading-log} \,.
\end{equation}
Compton scattering and annihilation/pair-creation processes do not give
equal contributions beyond leading log, but that does not concern
us in the present paper.

\section {Electrical conductivity}
\la {sec:conductivity}

To find the electrical conductivity, one must determine
the departure from equilibrium which is produced by an imposed electric field.
For the DC conductivity, one may take the electric field $\E$
to be constant and neglect all spatial and temporal variation
in the distribution function.
As discussed in section \ref {sec:kinetic},
only the external force
$
    {\bf F}^a_{\rm ext} = q^a \, \E
$
contributes on the left-hand side of the linearized Boltzmann equation
(\ref {eq:linearB}).
Rotational invariance implies that the correction $f_1^a(\p,x)$
to local equilibrium
may be taken to have the form (\ref {eq:fa1}) which, specialized
to the current case, reads
\begin {equation}
    f_1^a(\p,x)
    =
    -\beta^2 \, (\E \cdot \hat \p) \, \chi^a(p) \,.
\end {equation}
The linearized Boltzmann equation reduces to the form
$
    S_i^a (\p) = ({\cal C} \chi_i)^a(\p) \,,
$
with
$S_i^a(\p) \equiv -T q_a f_0^a(p) \, [1\pm f_0^a(p)] \, \hat p_i$
and
$\chi_i^a(\p) \equiv \hat p_i \, \chi^a(p)$.
The conductivity is given by
\begin {eqnarray}
    \sigma
    =
    \frac{\j \cdot \E}{\E^2}
    &=& 
    \frac{\E}{\E^2} \cdot \sum^\ffhc_{a} \> q_a
    \int_{\p} \hat\p \, f_0^a(p) \, [1 - f_0^a(p)] \, f_1^a(\p)
\nonumber \\ &=&
    -{\textstyle {1\over 3}}
    \beta^2 \sum^\ffhc_a \> q_a \int_\p f_0^a(p) \, [1 - f_0^a(p)] \, \chi^a(p)
\nonumber \\ &=&
    {\textstyle {2\over 3}}
    \left. Q_{\rm max} \right|_{\ell=1,q=q_{\rm EM}} \,,
\label {eq:sig}
\end{eqnarray}
where $Q_{\rm max}$ is the maximal value of the functional
$
    Q[\chi]
    =
    ( \chi_i, S_i )
    - \half \, ( \chi_i, {\cal C} \chi_i ) ,
$
first defined in Eq.~(\ref {eq:Q1}).
[In the explicit forms shown in Eq.~(\ref {eq:sig}),
we have specialized to only fermionic charge carriers, as is
appropriate for both pure QED and the full standard model.]

The driving term $S_i(\p)$ in the linearized Boltzmann equation
is charge conjugation (C) odd,
and therefore the departure from equilibrium $\chi(\p)$
must likewise be C odd.%
\footnote
    {%
    Alternatively, $S_i(\p)$ is CP even,
    so $\chi_i(\p) = \hat p_i \chi(p)$ is CP even,
    which means $\chi(p)$ itself is CP odd.
    }
Hence,
$\chi^{e^-}\!(p) = - \chie(p)$, while $\chi^\gamma(p)=0$.
In our high temperature regime,
the departure from equilibrium will also be identical for
different helicities of the same species,
and for different leptons of the same charge
if there are multiple light lepton species, so that
$\chi^{\mu^+}=\chie$, {\it etc}.
If there are active quark species, then rapid QCD scattering
processes (on the time scale of the relevant electromagnetic interactions)
ensure that $\chi^q(p)/\chie(p) = 0$ up to 
$
O[(\alpha_{\rm EM}/\alpha_s)^2]
$ 
corrections, which we will neglect.
(Active quarks remain relevant, however,
as targets from which the charged leptons can scatter.)
Hence, for this application we may regard $\chie(p)$ as the only
independent function which must be determined.

Let $N_{\rm leptons}$ denote the number of active lepton species
(so that $4 N_{\rm leptons}$ is the actual number of leptonic degrees of
freedom),
and let $N_{\rm species} = {1\over 4} {\sum\limits^{\tiny\ffhc}}_a (q_a/e)^2$
denote the 
sum over all Dirac fermion fields weighted by the square of their
electric charge assignments.
Then,
using the definition (\ref {eq:Q1}) of $Q[\chi]$ and the
leading-log forms (\ref {eq:B_is}) and (\ref {eq:D_is}) for the relevant
contributions to the collision integral,
the explicit form of the functional $Q[\chi]$ becomes
\begin {eqnarray}
    {Q[\chie \, ] \over N_{\rm leptons}}
    &=&
    -
    {2 \beta^2 e \over \pi^2}
    \int_0^\infty dp \> f_0^e(p) [1-f^e_0(p)] \>
    p^2 \chie(p)
\nonumber\\ && {}
    - (e^4 \ln e^{-1}) \,
    {N_{\rm species}\over 24 \pi^3} \,
    \int_0^\infty dp \> f_0^e(p) \, [1-f^e_0(p)] \>
    \left\{ p^2 \, [\chie(p)']^2 + 2 \, \chie(p)^2 \right\}
\nonumber\\ && {}
    - (e^4 \ln e^{-1}) \,
    {\beta \over 32 \pi^3}
    \int_0^\infty dp \> p \, f_0^e(p) \, [1+f^\gamma_0(p)] \> \chie(p)^2 \,.
\label {eq:Qcond}
\end {eqnarray}

Varying the above leading-log approximation to $Q[\chie \, ]$ with respect to
$\chie$ generates an ordinary differential equation for $\chie$, as was
noted in the context of shear viscosity by Heiselberg \cite{heiselberg}.
One option would be to solve that differential equation numerically.
However, we find it both numerically and conceptually more
convenient to instead directly attack the variational problem itself.
This is also good practice for going beyond leading-log order, where
the variational problem does not reduce to simple differential equations.

Since every term in this leading log form for $Q[\chie \, ]$ is a 
one dimensional integral over $p$, it is possible to perform the maximization
directly by discretizing values of $p$ and
then maximizing the resulting discrete quadratic form depending on a finite
set of values of $\chie(p)$.
This is completely straightforward, but requires a very fine discretization
in order to determine $Q_{\rm max}$ with high accuracy.
Another approach, which is
both efficient and remains practical when applied to the full leading
order in $g$ calculation,
is to maximize $Q[\chie \, ]$ within a variational subspace given by the
vector space spanned by a suitable set of basis functions,
$\{ \phi^{(m)}(p) \}$.
In other words, one uses an \ansatz
\begin{equation}
    \chie(p) = \sum_{m=1}^N a_m \, \phi^{(m)}(p) \, ,
\label {eq:basis-set}
\end{equation}
where $N$ is the size of the basis set considered,
and the coefficients $a_m$ are variational parameters
which will be tuned to maximize $Q[\chie \, ]$.
Inserting the \ansatz\ (\ref {eq:basis-set}) into
the functional (\ref {eq:Qcond}) produces an $N$-dimensional quadratic form,
\begin {equation}
    \tilde Q[\{a_m\}]
    =
    \sum_{m=1}^N
    a_m \, \tilde S_m
    -\half \sum_{m,n=1}^N
    a_m \, \tilde C_{mn} \, a_n \,,
\end {equation}
where the basis set components of the source vector,
$
    \tilde S_m \equiv \Big(S_i, \phi_i^{(m)} \Big)
$
and of the linearized collision operator,
$
    \tilde C_{mn} \equiv \Big(\phi_i^{(m)}, \, {\cal C} \phi_i^{(n)} \Big)
$
may be read off from the previous expression (\ref {eq:Qcond}).
Maximizing $\tilde Q$ is now a trivial linear algebra exercise
which gives
$
    a = \tilde C^{-1} \, \tilde S
$
and
\begin {equation}
    \sigma
    =
    {\textstyle {2\over3}} \, \tilde Q_{\rm max}
    =
    {\textstyle {1\over3}} \, a \cdot \tilde S
    =
    {\textstyle {1\over3}} \, \tilde S^\trans \, \tilde C^{-1} \, \tilde S
    \,,
\label {eq:sigmaa}
\end {equation}
where $a = ||a_m||$ and $\tilde S = ||\tilde S_m||$
are the $N$-component coefficient and source vectors, respectively,
in the chosen basis, and $\tilde C \equiv || \tilde C_{mn} ||$
is the (truncated) collision matrix.
Given a particular choice of basis functions, the individual components
$\{ \tilde S_m \}$ and $\{ \tilde C_{mn} \}$ may be computed by
numerical quadrature, and then a single $N \times N$ matrix inverse
yields the conductivity via the result (\ref {eq:sigmaa}).

It remains to choose a good family of basis functions for the variational
\ansatz.
There is a surprisingly good {\em single} function ``basis set,'' namely
$\phi^{(1)}(p) \equiv p/T$.
Its use permits one to find analytic expressions for
$\tilde S_1$ and $\tilde C_{11}$,
\begin{eqnarray}
    \tilde S_1^{\rm 1-parameter}
    &=&
    -N_{\rm leptons} \, \frac{9 \zeta(3)}{\pi^2} \> (eT) \,,
\\
    \tilde C_{11}^{\rm 1 - parameter}
    &=&
    N_{\rm leptons}
    \left( \frac{N_{\rm species}}{24 \pi} + \frac{\pi}{2^{8}} \right)
    \, (e^4 T \ln e^{-1}) \,.
\end{eqnarray}
When substituted into Eq.~(\ref{eq:sigmaa}), this yields
our previously quoted approximate result Eq.~(\ref{eq:approx-cond}).
This one parameter variational \ansatz\ is surprisingly accurate
for several reasons.
First, if one uses Boltzmann statistics instead of the correct
Bose or Fermi distributions,
[so that all appearances of $f_0(1{\pm}f_0)$ are replaced by $e^{-p/T}$],
then this one-parameter \ansatz\ turns out to be exact.
For large momenta, $p \gg T$,
this modification produces negligible change in the integrands
appearing in (\ref {eq:Qcond}),
but it seriously mangles the integrands
at small momenta, $p \ll T$.
However, the dominant contribution to the integrals in Eq.~(\ref {eq:Qcond})
comes from $p/T \sim 4$, where the effects of quantum statistics
are already rather small.
Additionally, 
at smaller momenta the Boltzmann approximation turns out to
over-estimate the contribution of diagram $(C)$ to the scattering
integral, but to under-estimate the contributions of diagrams $(D)$ and
$(E)$.  Hence there is some
cancellation when the contributions are comparable.
Finally, as in any variational approach,
the error in the resulting conductivity scales as the square
of the error in the function $\chie(p)$.
For these reasons, a well chosen one parameter \ansatz\ does
surprisingly well.

\begin{figure}[t]
\centerline{\epsfxsize=3.0in\epsfbox{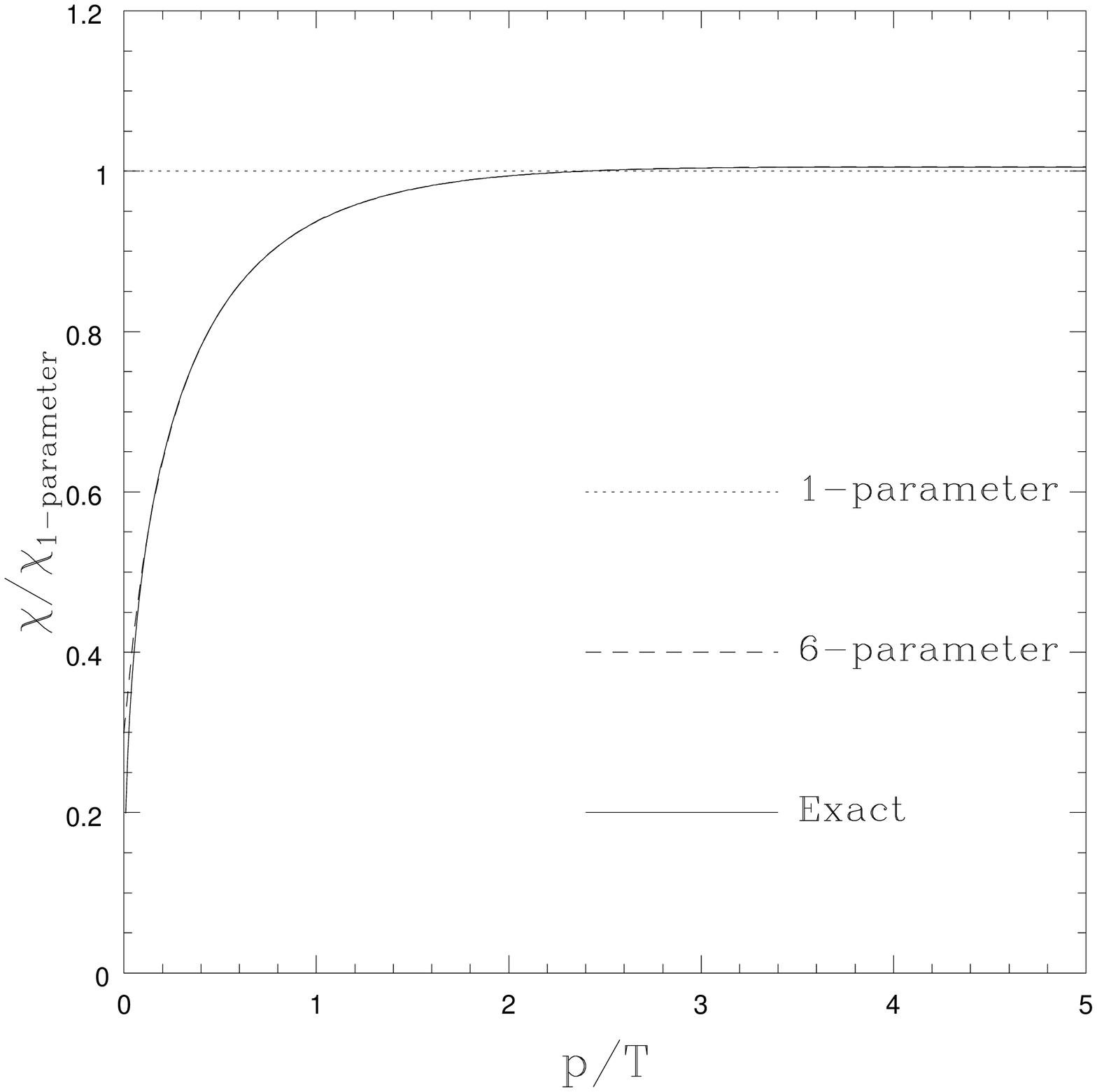} 
\hspace{0.1in} \epsfxsize=3.0in\epsfbox{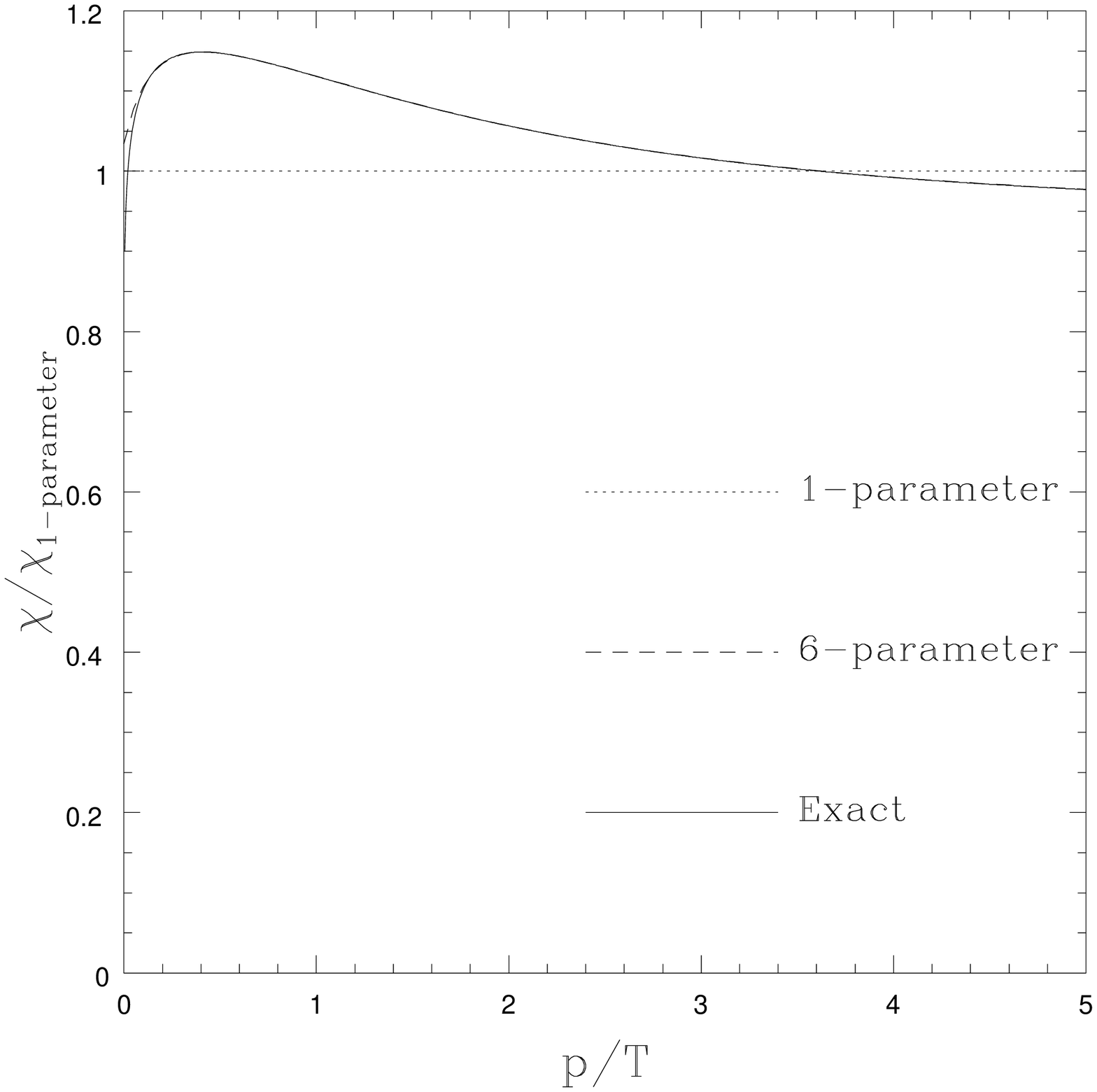} }
\vspace{0.1in}
\caption {%
    \la {fig:wave}%
    Comparison of the exact (leading-log) form for $\chi^e(p)$ to the
    one and six parameter \ansatz\ results,
    for the case of an $e^+ e^-$ plasma (left) and a three lepton 
    and five quark plasma (right).
    The six parameter \ansatz\ curve is essentially indistinguishable
    from the exact curve except for $p<0.1 T$.
    The dashed curves look exactly the same for both of our two \ansatze.
    }%
\end{figure}
We have investigated two larger basis sets,
each of which contains $\chi=p/T$ as a special case.%
\footnote
    {%
    The basis sets (\ref {eq:basis1}) and (\ref {eq:basis2})
    were actually selected in order to yield
    good results in our full leading-order calculations \cite {all-log}.
    For that application, it is helpful to have a non-orthogonal basis
    of strictly positive functions.
    For a leading-log analysis,
    many other simple choices of basis set would be equally good.
    }
One choice for an $N$-element basis set is
\begin{equation}
    \phi^{(m)}(p) = \frac{(p/T)^{m}}{(1+p/T)^{N-1}} \, , \quad
	m = 1,\ldots,N  \, .
\label {eq:basis1}
\end{equation}
The second is
\begin{equation}
    \phi^{(1)}(p) = \frac{p}{T} \,, \quad
    \phi^{(m)}(p) = \frac{p}{T} \, e^{-p/(c_m T)} \, , \quad
	m = 2,\ldots,N  \,,
\label {eq:basis2}
\end{equation}
with
\begin {equation}
    (c_2,c_3,c_4,\cdots)
    =
    ( \log\frac12,\, \log\frac34,\, \log\frac14,\, \log\frac78,\,
	\log\frac58 ,\, \log\frac38, \, \log\frac18, \, \cdots ) \, .
\end {equation}
For each basis set, we find that using the first
four basis functions determines the
(leading-log) value for $\sigma$ with a relative error less than $10^{-5}$,
while six basis functions are good to better than one part in $10^{6}$.
The determination of $\chie(p)$ itself is also excellent; 
we compare the ``exact'' function $\chie(p)$
[obtained from a discretization with very large $N$]
to the 1 and 6 parameter variational \ansatze\
in Fig.~\ref{fig:wave}.
The error barely visible in the six parameter \ansatz\ for
$\chie$ at $p \lsim T/10$ is quite irrelevant to the determination
of the conductivity because this region of the integration domain
contributes virtually nothing to the relevant integrals.
One may show that
the exact (leading-log) $\chie(p)$ function vanishes non-analytically
as $p\to0$ but this behavior has not been built into our
choice of basis sets.
Our numerical results for the leading-log conductivity
have already been presented in Table \ref{cond_table} of the summary.
They are based on six test functions,
and are accurate to the last digit shown.  

The best previous result in the literature for the leading-log
electric conductivity in a relativistic plasma is that of
Baym and Heiselberg \cite{BaymHeiselberg}.
Their analysis differs from ours in the following respects.
First, they neglect diagrams $(D)$ and $(E)$, and the
same-charge contributions (particle on particle or anti-particle on
anti-particle) in diagram $(C)$.%
\footnote{%
   They argue that $e^-e^-\to e^-e^-$ and $e^+e^+\to e^+e^+$ processes
   do not change the net current and so do not affect the
   conductivity.  The final inference is incorrect except in the
   one-parameter ansatz.  Though these collisions do not change the
   net current, they do change the distribution of velocities, which can
   then indirectly affect the rate at which the current is changed
   by other processes (although numerically the effect is quite small).
   Specifically within the one-parameter ansatz, however,
   the $e^+$ and $e^-$ distributions are individually just normal
   thermal equilibrium distributions boosted parallel and anti-parallel
   to the electric field; and so for this restricted ansatz
   these scattering processes have no effect at all.
}
Second, they consider only the one parameter \ansatz, $\chi(p)=p/T$,
and they determine its magnitude by
integrating both sides of the Boltzmann equation against $\hat\p$,
rather than against $\hat\p \chi(p)$.
Their treatment is thus not a variational approximation,
and consequently the result for the conductivity is linearly,
not quadratically, sensitive to the error in the $\chi(p)$.
If we drop the contribution of diagrams $(D)$ and $(E)$ from our analysis,
our result for $\sigma$ is $7.6\%$ lower than theirs;
including all diagrams, for a pure $e^+ e^-$ plasma,
our result for $\sigma$ is just under half of theirs.

\section {Flavor diffusion}
\la {sec:diffusion}

As discussed in section \ref {sec:kinetic},
the determination of flavor diffusion constants exactly parallels
the calculation of conductivity, except for the use of differing
charge assignments appropriate for the globally conserved current of interest,
and an overall factor of the corresponding charge susceptibility
$\suscept_\alpha \equiv \partial n_\alpha / \partial \mu_\alpha$.

We will consider a theory with a simple gauge group,
some set of active matter fields (fermions and/or scalars)
in arbitrary representations of the gauge group,
and no other significant interactions.
Hence, the net number density of each matter species (or `flavor')
is a conserved charge which will relax diffusively.
We will only consider flavors carried exclusively by fermions.
The diffusion constant for the net number density of species $a$ is,
from Eq.~(\ref {eq:DQ}), given by
\begin {equation}
    D_a
    =
    {\textstyle {2\over3}} \, Q_{\rm max}\Big|_{\ell=1,q=q_a}
    \Bigm/ \suscept_a \,,
\end {equation}
where the functional $Q[\chi]$ is to be maximized in the
$\ell=1$ channel with charge $\pm 1$ assigned to all particles or antiparticles
of species $a$, respectively, and 0 to all other excitations.
To be consistent, all particles and anti-particles of species $a$ should
then also be
included in the number density $n^a$ used to define the susceptibility
$\suscept_a$ [from Eq.~(\ref {eq:dndmu})], giving
\begin {equation}
    \suscept_a = {\partial n_a \over \partial \mu_a}
    =
    {\textstyle {1\over 3}}\,
    d_{{\rm R}_a}
    \, {T^2}\,.
\end {equation}
Once again, charge conjugation (or CP) invariance implies that
the particle and antiparticle departures from equilibrium are equal
and opposite, $\chi^b = -\chi^{\overline b}$, while the
gauge boson departure from equilibrium must vanish, $\chi^g = 0$.%
\footnote
    {%
    This follows because the gauge boson distribution,
    which comes from the color-singlet part of its density matrix,
    is charge conjugation even,
    while the conserved charge of interest is charge conjugation odd.%
    }

In the leading-log approximation,
the gauge boson exchange contributions (\ref {eq:B_is})
to the collision operator are diagonal in particle species.
[That is, there are no $\chi^a \chi^b$ cross terms in (\ref {eq:B_is}).]
And when symmetry prevents the gauge bosons from having any
departure from equilibrium,
the leading-log annihilation/Compton scattering contributions
(\ref {eq:D_is}) to the collision operator are also diagonal in
particle species.
Consequently, in the case at hand,
where the driving term $S_i$ in the linearized
Boltzmann equation is only non-zero for species $a$
and $\chi^g$ vanishes,
the only non-zero departure from equilibrium emerging
from the linearized Boltzmann equation (or the maximization of $Q[\chi]$)
will be for species $a$.

More physically, the essential point is that in the theory under
consideration, in leading-log approximation,
a departure from equilibrium in the net number density for species $a$
will relax diffusively {\em without\/} inducing a net number
density in any other species.
Thus, for the determination of the diffusion constant $D_a$,
the variational functional $Q[\chi]$ may be expressed solely
in terms of $\chi^a$.
Using the leading-log forms (\ref {eq:B_is}) and (\ref {eq:D_is})
for the relevant contributions to the collision integral, and
inserting the species $a$ particle number as the relevant charge in the
source term (\ref {eq:Sij}),
the explicit form of the functional $Q[\chi]$ becomes
\begin {eqnarray}
    {Q[\chi^a] \over d_{{\rm R}_a}}
    &=&
    -
    {2 \beta^2 \over \pi^2}
    \int_0^\infty dp \> f_0^a(p) [1 - f^a_0(p)] \>
    p^2 \chi^a(p)
\nonumber\\ && {}
    - (g^4 \ln g^{-1}) \,
    {C_{{\rm R}_a} \over 96\pi^3}
    \Biggl[ \sum^\ffh_b T_{{\rm R}_b} \lambda_b \Biggr]
    \int_0^\infty dp \> f_0^a(p) \, [1 - f^a_0(p)] \>
    \left\{ p^2 \, [\chi^a(p)']^2 + 2 \, \chi^a(p)^2 \right\}
\nonumber\\ && {}
    - (g^4 \ln g^{-1}) \,
    {C_{{\rm R}_a}^2 \beta \over 32\pi^3}
    \int_0^\infty dp \> p \, f_0^a(p) \, [1+f^g_0(p)] \> \chi^a(p)^2 \,.
\label {eq:Qdiffuse}
\end {eqnarray}
Inserting a finite basis-set expansion for $\chi^a(p)$,
as in Eq.~(\ref {eq:basis-set}),
extracting the resulting basis set components
for the source vector $\{ \tilde S_m \}$ and collision matrix
$\{ \tilde C_{mn} \}$, and inverting the (truncated) collision matrix
to determine $\tilde Q_{\rm max}$,
proceeds exactly as described in the previous section.

Using the same one-term \ansatz\ $\chi^a \propto p/T$
with the functional (\ref {eq:Qdiffuse}) leads to our
previously quoted analytic approximation (\ref{eq:approx_D}).
Specialized to an $SU(3)$ gauge theory with $\nf$ fundamental representation
quarks, this reduces to Eq.~(\ref{eq:D_F}).
Maximizing the above $Q[\chi^a]$ using six or more terms
of either basis set (\ref {eq:basis1}) or (\ref {eq:basis2}),
and the specific values $\cf=4/3$, $\tf = 1/2$,
and $\ta=3$ appropriate for SU(3),
leads to the numerical values shown in Table~\ref{diff_table}.

Four previous determinations of the QCD diffusion coefficient merit mention.
The first is by Heiselberg \cite{Heiselberg_diff}.
Within the single parameter \ansatz, $\chi^a(p) \propto p/T$,
his analysis coincides with ours,
except that he leaves out diagrams $(D)$ and $(E)$.
He makes an additional factor of $16/9$ error in his final value for the
diffusion constant,
which is therefore just more than twice ours for the case $\nf=6$.
We are unable to trace the origin of this error;
equations (9)--(12) of his paper appear correct, but (13) is not.
Joyce, Prokopec, and Turok \cite{JPT1} also treat quark diffusion.
In this paper they make a leading log treatment in which they only
consider diagram $(C)$, which means they only keep the $\nf \tf$
part of the second term in Eq.~(\ref{eq:Qdiffuse}).
Furthermore, they neglect the energy transfer of a collision,
which effectively means that they drop the $p^2 [\chi^a(p)']^2$
term in the associated integral but keep the $2[\chi^a(p)]^2$ term.
Under this ``approximation,'' it is not difficult to solve exactly for the
resulting form for $\chi(p)$ without any recourse to an \ansatz.
Their result for the diffusion constant is about $5\%$ higher than our result
would be if we only included diagram $(C)$
[which means retaining only the middle term in the denominator
of Eq.~(\ref{eq:D_F})].
In fact diagram $(C)$ is subdominant, so their result is fairly far off;
for a six quark plasma their result is 2.3 times ours,
and for fewer quarks it is worse.
They also treat lepton diffusion, making the same approximation
and a corresponding sized error.
Another treatment of the baryon number diffusion constant is
given by Moore and Prokopec \cite{MooreProkopec}.
Their analysis coincides with our treatment if we make the one parameter 
\ansatz\ $\chi^a(p) \propto p/T$.
They did include diagrams $(B)$ through $(E)$.
However, they failed to include the (1/2) symmetry factor in
diagram $(D)$ and make an algebraic error in diagram $(E)$, so their
results for these diagrams are off by almost a factor of 2,
and their result for the resulting diffusion constant is about 12\%
smaller than ours.
The treatment in another paper of Joyce, Prokopec, and Turok \cite{JPT2} is
similar to the treatment of Moore and Prokopec, but only includes
diagram $(C)$ and makes some errors in its evaluation.

\section {Shear viscosity}
\la {sec:shear}

As shown in section \ref {sec:kinetic}, the solution of the linearized
Boltzmann equation needed to determine the shear viscosity is equivalent
to the maximization of the functional
\begin{equation}
    Q[\chi]
    =
    \Big( S_{ij} , \chi_{ij} \Big)
    - \half \Big( \chi_{ij} , {\cal C} \chi_{ij} \Big) \, ,
\label {eq:Qshear0}
\end{equation}
now specialized to $\ell = 2$,
with $S_{ij}^a(\p) = -T f_0^a(p) [1\pm f_0^a(p)] \, |\p| \, I_{ij}(\hat \p)$,
$\chi_{ij}^a(\p) = I_{ij}(\hat \p) \, \chi^a(p)$,
$
    I_{ij}(\hat \p) =
    \sqrt{{3\over 2}} \,
    \left( \hat p_i \hat p_j - {1\over3}\delta_{ij} \right)
$,
and the linearized collision operator $\cal C$ defined by Eq.~(\ref {eq:Q2}).
The resulting viscosity is proportional to the maximal value of $Q[\chi]$,
\begin {equation}
    \eta = {\textstyle {2 \over 15}} \> Q_{\rm max} \,.
\end {equation}

The source term $S_{ij}$ is invariant under all flavor symmetries
as well as charge conjugation and CP.
Hence, the departures from equilibrium which solve the linearized
Boltzmann equation $S_{ij} = {\cal C} \chi_{ij}$
(or equivalently, which maximize $Q[\chi]$) will also be invariant
under these symmetries.
So, for example, in QCD, every quark and anti-quark species
will have the same departure from equilibrium
which we will denote as $\chi^q$,
and all gluons will share another common departure $\chi^g$.
But there is no reason for $\chi^q$ to equal $\chi^g$,
nor for either one to vanish.
Hence, two distinct functions
must be varied independently to find the maximum of $Q[\chi]$.

Using the leading-log forms (\ref {eq:B_is}) and (\ref {eq:D_is})
for the relevant contributions to the collision integral,
and specializing to a QCD-like theory,
the two terms in $Q[\chi^g,\chi^q]$, Eq.~(\ref {eq:Qshear0}),
explicitly equal
\begin {equation}
\Big( S_\ij , \chi_\ij \Big)
    =
    -{\beta^2 \over \pi^2}
    \int_0^\infty \! dp \> p^3 \>
    \Bigl\{
	\da \, f_0^g(p) [1 {+} f^g_0(p)] \, \chi^g(p)
	+
	2\, \df\nf \, f_0^q(p) [1 {-} f^q_0(p)] \, \chi^q(p)
    \Bigr\} \,,
\label {eq:Sshear}
\end {equation}
and
\begin {eqnarray}
{\Big( \chi_\ij , {\cal C} \chi_\ij \Big) \over g^4 \ln g^{-1}}
    &=&
    {\da \ta \over 24 \pi^3} \,
    ( \ta {+} \nf\tf )
    \int_0^\infty dp \>
	f_0^g(p) \, [1{+}f^g_0(p)] \>
	\left\{ p^2 \, [\chi^g(p)']^2 + 6 \, \chi^g(p)^2 \right\}
\nonumber\\ &+&
    {\nf \da \tf \over 12 \pi^3} \,
    ( \ta {+} \nf \tf )
    \int_0^\infty dp \>
	f_0^q(p) \, [1 {-} f^q_0(p)] \>
	\left\{ p^2 \, [\chi^q(p)']^2 + 6 \, \chi^q(p)^2 \right\}
\nonumber\\ &+&
    {\nf\da\tf\cf \beta \over 16\pi^3}
    \int_0^\infty dp \> p \, f_0^q(p) \, [1{+}f^g_0(p)] \>
	\Big[ \chi^q(p) - \chi^g(p) \Big]^2 \,.
\label {eq:Cshear}
\end {eqnarray}

Our approach for maximizing $Q[\chi]$
is an obvious generalization of the previous treatment used
for conductivity or diffusion.
To carry out the maximization within a finite dimensional variational
subspace, we expand each of the undetermined functions in
a finite basis set,
\begin{equation}
    \chi^g(p) = \sum_{m=1}^N a_m \, \phi^{(m)}(p) \, , \qquad
    \chi^q(p) = \sum_{m=1}^N a_{N+m} \, \phi^{(m)}(p) \, ,
\label{eq:chi_visc}
\end{equation}
with coefficients $\{a_m\}$, $m=1,\ldots, 2N$, that are independent
variational parameters.
The variational functions $\phi^{(m)}(p)$ we use for shear viscosity
are $p$ times those presented in Eq.~(\ref{eq:basis1}) or (\ref{eq:basis2}),
because the conserved charge in question is now proportional to $p$.
Inserting Eq.~(\ref{eq:chi_visc}) into expressions
(\ref {eq:Sshear}) and (\ref {eq:Cshear}),
one may read off the basis-set components of the source vector $\tilde S$
and truncated scattering matrix $\tilde C$,
\begin{equation}
    \Big( S_{ij} , \chi_{ij} \Big) = \sum_m a_m \,\tilde S_m \, , \qquad
    \Big( \chi_{ij} , {\cal C} \chi_{ij} \Big)
    = \sum_{m,n} a_m \,\tilde C_{mn} \, a_n \, .
\end{equation}
As before, one has
$\tilde Q[\{a_m\}] = a^\trans \tilde S - \half a^\trans \tilde C \, a$.
At the maximum, the vector of coefficients $a = \tilde C^{-1} \tilde S$,
and
\begin {equation}
    \eta = {\textstyle {2\over 15}} \, \tilde Q_{\rm max}
    = {\textstyle {1\over 15}} \, a \cdot \tilde S
    = {\textstyle {1\over 15}} \, \tilde S^\trans \tilde C^{-1} \tilde S \,.
\label {eq:etatilde}
\end {equation}
The only change from the previous cases is that a set of $N$ basis functions
$\{ \phi^{(m)} \}$ now generates a $2N$ dimensional linear algebra problem.
Note that non-zero block off-diagonal components of $\tilde C_{mn}$
(those with $m \le N < n$ or $m>N\ge n$)
represent a cross-coupling between the quarks and gauge bosons.
These matrix elements arise only from the last term in Eq.~(\ref {eq:Cshear}),
which was generated by pair annihilation and Compton scattering processes
[diagrams $(D)$ and $(E)$].
If those contributions were absent, then $Q$ would split
into separate boson and fermion pieces, and the viscosity
would be a sum of two independent terms, one due to fermions and the other
due to the gauge bosons.

The natural one-function \ansatz\ is now $\phi^{(1)}(p) = p^2/T$.
Once again, this exactly solves the Boltzmann equation if the
$f_0 [1\pm f_0]$ quantum statistics factors are replaced by classical
Boltzmann statistics.
With this simple \ansatz, one may evaluate the integrals in
(\ref {eq:Sshear}) and (\ref {eq:Cshear}) analytically.
One finds
\begin{equation}
    \tilde S
    =
    -{120 \, \zeta(5) \, T^3 \over \pi^2}
    \left[
	\begin {array}{c}
	    \da
	    \\
	    {15 \over 8} \, \df\nf
	\end {array}
    \right],
\end{equation}
and
\begin{eqnarray}
    \tilde C
    &=&
    {\pi T^3 \over 9 \, \da} \,
    \Biggl(
    (\da\ca+\nf\df\cf) \,
    \left[
	\begin {array}{cc} \da\ca & 0 \\ 0 & {7\over4} \, \nf\df\cf
	\end {array}
    \right]
     +
     {9\pi^2\over 128} \,\nf\df\,\cf^2\,\da
	\left[ \begin {array}{rr} 1 & -1 \\ -1 & 1 \end{array} \right]
    \Biggr)
\nonumber\\ && {}
    \times (g^4 \ln g^{-1})
    \,.
\label {eq:Ctilde}
\end{eqnarray}
Except for the numerical prefactors (and $g^4 \ln g^{-1}$), these
are exactly expressions (\ref {eq:v}) and (\ref {eq:c}) given in
the Introduction.
When combined with the relation (\ref {eq:etatilde}), these
one-term \ansatz\ results yield the approximate form (\ref {eq:etaapprox})
quoted earlier.

Maximizing $Q[\chi^g,\chi^q]$ for an $SU(3)$ theory,
using six or more terms of a basis set with the form of
either (\ref{eq:basis1}) or (\ref {eq:basis2}),
but with each basis function multiplied by one additional factor of $p$,
leads to the numerical values shown in Table \ref{shear_table}.

It should be noted that
the role played by the final term of (\ref {eq:Cshear})
is {\em not} to relax the 
traceless part of the stress tensor (or momentum flux) of the plasma.  
At leading log order, the traceless stress is the same before and after an
annihilation (or Compton scattering) process,
because the outgoing momenta of the gauge bosons approximately
equal the incoming momenta of the fermions.
This is why the sum of these contributions vanish if $\chi^g = \chi^q$.
What these processes do is to {\em transfer} momentum between the
fermions and the bosons.
Such a transfer is important,
because the gauge bosons and fermions equilibrate at different rates.
The gauge bosons, with their larger gauge group Casimir,
equilibrate faster than the fermions.
Thus, the presence of a channel which transfers a departure
from equilibrium in the slowly relaxing fermions,
into a departure in the more rapidly relaxing bosons,
speeds the relaxation process.

This is why our result for $\eta$ is
somewhat lower than that found by Baym {\em et~al.}~\cite{BMPRa}, who
missed diagrams $(D)$ and $(E)$, but whose treatment otherwise coincides 
with ours when we use the one term \ansatz.
For the case of three color QCD,
the difference is largest at $\nf = 2$, where it is a $3\%$ effect.
Omitting these diagrams is significantly less important for viscosity
than it is for conductivity and diffusion.  This is partly because
scattering via gauge boson exchange becomes more efficient at higher $\ell$
due to the $\ell(\ell{+}1)$ factor in Eq.~(\ref{eq:B_is}),
and partly because the processes of diagrams $(D)$
and $(E)$ only indirectly affect the relaxation in this case,
whereas for conductivity or
diffusion the annihilation and Compton scattering processes
can directly relax the relevant flux.

Abelian gauge theories are a special exception.
In a QED-like theory,
diagrams $(A)$ and $(B)$ are absent (because $\ca=0$),
and as a result diagrams $(D)$ and $(E)$ are essential.
Without these diagrams, photon fluctuations are not damped by any
logarithmically enhanced process.  If one omits
these processes, then the matrix $\tilde C$ becomes singular and the
resulting leading-log viscosity diverges.
This also helps explain why the QED viscosity we find in
Eq.~(\ref{eq:visc_QED}) is so large.

For a more complicated theory with a product gauge group,
such as QCD plus QED,
or the full standard model at temperatures above $M_W$,
it is necessary to treat the departure from equilibrium of each
species with differing gauge couplings as a distinct variable;
so there are many independent $\chi$'s.
There are also a large number of diagrams.
However, if one of the gauge couplings is much smaller than the others,
so that corrections of order of the fourth power of the ratio of couplings
are negligible, then one may substantially simplify the calculation.
This criterion certainly holds for QCD plus QED,
where $\alphaEM^2 \ll \alphas^2$, and it is a
reasonable approximation for the standard model,
where $(g'/\gw)^4 < 0.1$.
In this situation, all degrees of
freedom which couple to the stronger gauge group, and all gauge bosons,
may be regarded as remaining in equilibrium.
Hence only weakly coupled fermions --- leptons for QCD plus QED,
right handed leptons for the high temperature standard model ---
are out of equilibrium.  
This approximation is valid for the more strongly interacting particles
because their relaxation rates are much larger.
In QCD plus QED, for example,
a quark's departure from equilibrium relaxes
at a rate which is $O[(\alphas/\alphaEM)^2]$ faster than the
corresponding rate for the electron.
This approximation is also valid for the more weakly interacting
gauge bosons, such as the photon,
because of the mixed weak/strong scattering processes represented by
diagrams $(D)$, $(E)$, and $(J)$,
when the fermion is a quark, one gauge line is
a gluon, and the other is a photon.
This process occurs at a rate of order
$e^2 \gs^2 T$, which is fast compared to the
$O(e^4 T)$ (up to logs) relaxation rate of leptons.
Similarly, in the hot standard model, 
hyper-photons scatter at a rate of order $g'{}^2 \gs^2 T$
which is $O(\alphas/\alpha')$ larger than the large angle
scattering rates for right-handed leptons.
This is the approximation we have used to
obtain the viscosity results
discussed in Sec.~\ref{sec:intro}
for QCD plus QED, and for the standard model at
high temperature.

The shear viscosity in gauge theories
has previously been considered by
Baym, Monien, Pethick, and Ravenhall \cite{BMPRa}.
Their treatment is quite close to ours, and
correctly analyzes the scattering contributions of diagrams $(A)$--$(C)$ 
within the single function \ansatz\ $\chi \propto p^2$.
They also explore an \ansatz\ with $\chi \propto p^\alpha$
in an effort to test the quality of the $p^2$ \ansatz.
However, they miss diagrams $(D)$ and $(E)$
which, as noted above, means that their values for $\eta$ are
slightly too large.  Shear viscosity has also been considered by Heiselberg
\cite{heiselberg}.  For the case of pure glue ($\nf=0$), Heiselberg
gives a complete treatment, solving the variational problem for $\chi$
without recourse to any \ansatz.  We agree with his result for the $\nf=0$
viscosity, except for a trivial sign error in his presentation of the
difference with respect to the $p^2$ \ansatz\ value.  However his treatment
of the nonzero $\nf$ case contains some errors not present in \cite{BMPRa}, 
apparently from an incorrect treatment of statistical factors in diagram $(B)$.

\section* {ACKNOWLEDGMENTS}

This work was supported, in part, by the U.S. Department
of Energy under Grant Nos.~DE-FG03-96ER40956
and DE-FG02-97ER41027.

\appendix

\section{Group factor notation}
\label{app:group}

For any representation R of a simple gauge group,
let $\{ t^a \}$ denote the representation matrices of the
Lie algebra generators.
The normalization $T_{\rm R}$ of that representation is
defined by
\begin {equation}
   {\rm tr} \> t^a \, t^b \equiv T_{\rm R} \, \delta^{ab} .
\end {equation}
The quadratic Casimir $C_{\rm R}$ is defined by
\begin {equation}
   \sum_a (t^a)^2 \equiv C_{\rm R} \, {\bf 1} .
\end {equation}
The two are related by
\begin {equation}
   T_{\rm R} \, d_G = d_{\rm R} \, C_{\rm R} ,
\end {equation}
where $d_{\rm R}$ is the dimension of the representation R
and $d_G{=}d_{\rm A}$ is the dimension of the gauge group.
For the fundamental representation of $SU(N)$,
\begin {equation}
   \cf = {N^2{-}1 \over 2N},
   \qquad
   T_{\rm F} = \half,
   \qquad
   \df = N,
\end {equation}
while for the adjoint representation,
\begin {equation}
   \ca = N,
   \qquad
   \ta = N,
   \qquad
   \da = N^2-1 .
\end {equation}

The above notation is most natural to non-Abelian theories.  However, results
in the text can be translated to Abelian theories by taking $d_A = 1$.
Then, for each species $c$ of charged particles, take $d_R=1$ and identify
the generator $t$ with the charge assignment $(e_c/e)$ of that species.
The group factors are then
\begin {equation}
   C_{\rm R_c} = T_{\rm R_c} = e_c^2/e^2,
   \qquad
   \ca = \ta = 0 ,
   \qquad
   \df = 1 .
\end {equation}

\end {document}